\documentclass[12pt]{iopart}

\pdfoutput=1

\usepackage{graphicx,sidecap}
\usepackage{bm}
\usepackage{braket}
\usepackage{amssymb}
\usepackage{mathrsfs}
\usepackage{amsmath}
\usepackage{xcolor}
\usepackage{hyperref}
\usepackage{enumerate}
\usepackage[toc,title]{appendix}

\hypersetup{colorlinks,bookmarksopen,bookmarksnumbered,
citecolor=red,
linkcolor=blue,
pdfstartview=green,
urlcolor=purple}

\catcode`@=11 
\renewcommand\tableofcontents{%
  \section*{\contentsname}%
  \@starttoc{toc}%
}
\catcode`@=12

\def\be{\begin{equation}}
\def\ee{\end{equation}}

\def\bea{\begin{eqnarray}}
\def\eea{\end{eqnarray}}

\def\Tr{{\rm Tr}}

\usepackage{xparse}

\DeclareDocumentCommand{\TrProd}{ m O{} o O{} o O{} o O{} o }{%
\{ \Gamma_{#1}^{#2}
\IfNoValueTF{#3}{}{,\Gamma_{#3}^{#4}}
\IfNoValueTF{#5}{}{,\Gamma_{#5}^{#6}}
\IfNoValueTF{#7}{}{,\Gamma_{#7}^{#8}}
\IfNoValueTF{#9}{}{,\Gamma_{#9}}
\}
}

\DeclareDocumentCommand{\TrProdTilde}{ m O{} o O{} o O{} o O{} o }{%
\{ \widetilde\Gamma_{#1}^{#2}
\IfNoValueTF{#3}{}{,\widetilde\Gamma_{#3}^{#4}}
\IfNoValueTF{#5}{}{,\widetilde\Gamma_{#5}^{#6}}
\IfNoValueTF{#7}{}{,\widetilde\Gamma_{#7}^{#8}}
\IfNoValueTF{#9}{}{,\widetilde\Gamma_{#9}}
\}
}

\DeclareDocumentCommand{\ch}{ m o o o m o o o }{%
\begin{bmatrix}
#1 \IfNoValueTF{#2}{}{& #2}\IfNoValueTF{#3}{}{& #3}\IfNoValueTF{#4}{}{& #4} \\
#5 \IfNoValueTF{#6}{}{& #6}\IfNoValueTF{#7}{}{& #7}\IfNoValueTF{#8}{}{& #8}
\end{bmatrix}_{\tau}
}

\newcommand{\conj}[1]{#1^{\ast}}
\newcommand{\intf}[1]{\int\text{d}\conj{#1}\text{d}#1\,}
\newcommand{\df}[1]{\text{d}\conj{#1}\text{d}#1\,}
\newcommand{\iu}{\text{i}}
\makeatletter
\newcommand{\pushright}[1]{\ifmeasuring@#1\else\omit\hfill$\displaystyle#1$\fi\ignorespaces}
\newcommand{\pushleft}[1]{\ifmeasuring@#1\else\omit$\displaystyle#1$\hfill\fi\ignorespaces}
\makeatother

\begin{document}

\title[
Towards entanglement negativity for a 1D free fermion
]{
\\
Towards entanglement negativity of two disjoint intervals  for a  one dimensional free fermion
}

\vspace{.5cm}

\author{Andrea Coser, Erik Tonni and Pasquale Calabrese}
\address{SISSA and INFN, via Bonomea 265, 34136 Trieste, Italy. }

\vspace{.5cm}

\begin{abstract}
We study the moments of the partial transpose of the reduced density matrix of two intervals for the free massless 
Dirac fermion. 
By means of a direct calculation based on coherent state path integral, we find an analytic form for these moments 
in terms of the Riemann theta function. 
We show  that the moments of arbitrary order are equal to the same quantities for the compactified boson at the self-dual point.  
These equalities imply the non trivial result that also the negativity of the free fermion and the self-dual boson are equal.
\end{abstract}

\maketitle

\tableofcontents

\section{Introduction}

The study of the entanglement content of extended quantum systems became in recent times a subject of 
extremely large theoretical interest (see e.g.\ the references in \cite{rev} as reviews).
While the bipartite entanglement for an extended system in a pure state is a well understood subject and it can be
quantified by the so-called entanglement entropy (i.e.\ the von Neumann entropy of the reduced density matrix of one of the two parts),
for a mixed state 
the quantification of the entanglement is a more subtle issue. 
The observation that the entanglement in a bipartite mixed state 
is related to the presence of negative eigenvalues in the partial transpose of the density matrix \cite{partial}
has led to the introduction of the negativity \cite{vw-02} which subsequently has been shown to be 
an entanglement monotone \cite{neg-mon}, i.e.\ a good entanglement measure from a quantum information point of view.


Although the negativity is a ``computable measure of entanglement" \cite{vw-02}, its direct and explicit computation in 
a many body system is very cumbersome. 
This difficulty can be (at least partially) overcome by a replica approach based on the computation of the even 
moments of the partially transpose density matrix \cite{cct-neg-letter}.
This recent approach has been already applied to the study of one-dimensional conformal field theories (CFT) 
in the ground state \cite{cct-neg-letter,cct-neg-long,ctt-13,rr-14}, in thermal state \cite{cct-neg-T,ez-14}, and in 
non-equilibrium protocols \cite{ez-14,ctc-14,hd-15,wcr-15}, as well as to topological systems \cite{c-13,lv-13,kor}.
Focusing on 1D CFTs in the ground state, the negativity is explicitly known only for the  simple (but non trivial) geometry of two 
adjacent intervals embedded in a larger system. For the very important case of the entanglement between two disjoint intervals 
only the limit of close intervals is explicitly known. 
For arbitrary distances between the intervals, the main difficulty is to find the analytic continuation of the even-integer moments to $n\to1$, 
although these moments are analytically known in a few cases \cite{cct-neg-long,ctt-13}  
(and indeed numerical interpolation techniques \cite{ahjk-14} have been exploited to have a numerical prediction for the negativity \cite{dct-15}). 
These analytical studies have been paralleled by several numerical works such as in Refs.
\cite{a-13,AlbaLauchlin-neg,wmb-09,aw-08,fcga-08,Neg3,sod}.

The goal of this manuscript is to investigate the negativity for the one-dimensional CFT of a free fermionic model. 
The main result is a close analytical form for the moments of the partial transpose of two disjoint intervals 
for the massless free Dirac fermion reported in Eq.~\eqref{eq:partial transpose free fermion}.
The explicit form of the moments is exactly the same as for the compactified boson at the 
self-dual point obtained in Ref.~\cite{cct-neg-long}.

The manuscript is organised as follows. 
In Sec.~\ref{sec: fermionic rhoA} we build the partial transpose of the fermionic density matrix using coherent state path integral. 
In Sec.~\ref{sec:FF} we provide the analytical form for the moments of the partial transpose of two disjoint intervals and we analyse it.  
Finally in Sec.~\ref{concl} we draw our conclusions and discuss some open problems. 
In a series of four appendices we report a number of technical details.

\section{Partial transpose of the reduced density matrix for the free fermion}
\label{sec: fermionic rhoA}

In this section we provide a path integral formula for the partial transpose of the density matrix for a free fermionic field theory, 
after a brief review of the result of Eisler and Zimboras \cite{ez-15} for the partial transpose of the reduced density 
matrix of two disjoint blocks on the lattice.

\subsection{Review of the lattice results}
\label{subsec lattice ff}

We start from the the tight binding model  with Hamiltonian 
\begin{equation} 
\label{eq:Ham}
H=\frac12 \sum_{i=1}^L \Big[c^\dagger_i c_{i+1}+ c^\dagger_{i+1} c_{i} \Big],
\end{equation}
where periodic boundary conditions are assumed. We only consider the model at half filling $k_F=\pi/2$.
Since (\ref{eq:Ham}) is quadratic in the fermionic operators, it can be diagonalized in momentum space. 
The scaling limit of this model is the massless Dirac fermion which is a CFT with $c=1$.
The local Hilbert space $\mathcal{H}_j$ of a single site is two dimensional and we can choose a basis made by the two 
vectors corresponding to whether the fermion occurs ($\ket{1}$) or not ($\ket{0}$).
In this basis the operators $c_j$ and $c_j^\dag$ act as the creation and annihilation operators, $c_j\ket{0}=c_j^\dagger \ket{1}=0$, 
while $c_j^\dagger \ket{0}= \ket{1}$ and $c_j \ket{1}= \ket{0}$.
The tight-binding model can be mapped by a Jordan-Wigner transformation into the XX spin chain 
\be
H_{XX}=\sum_{j=1}^L \Big[\sigma^x_{j}\sigma^x_{j+1}+\sigma^y_{j}\sigma^y_{j+1}\Big] .
\label{HXX}
\ee
Although these models are mapped one into the other, since the Jordan-Wigner transformation between them  is not local, the entanglement (both entropy and negativity) of two disjoint blocks are not equal, as pointed out already in the 
literature \cite{ip-10,atc-10}.

We always consider the entire system to be in the ground state $\ket{\Psi}$ with density matrix 
$\rho=\ket{\Psi}\bra{\Psi}$.
%
It is useful to introduce the following Majorana fermions \cite{vlrk-03} 
\begin{equation}
\label{a from c def}
a_{j}^x = c_j + c_j^\dagger ,
\qquad
a_{j}^y = \iu (c_j - c_j^\dagger) ,
\end{equation}
which satisfy the anticommutation relations $\{a^\alpha_r, a^\beta_s\}=2\delta_{\alpha\beta}\delta_{rs}$.
The single site Majorana operators can be also written  as
\begin{equation}
\label{majorana op}
 a_j^x = \Sigma_j^x ,  \qquad a_j^y = -\Sigma_j^y ,  \qquad \iu \,a_j^xa_j^y = \Sigma_j^z .
\end{equation}
The operators $\Sigma_j^\alpha$ on the single site satisfy the algebra of the Pauli matrices,
but at different sites they anticommute and so they are not proper spin operators and should not be 
confused with the $\sigma_j^\alpha$ in (\ref{HXX}). 
For each site, we also need to define the following unitary operator 
\begin{equation}
\label{Uk def}
 U_{\alpha}^{(k)} = e^{\iu\frac{\alpha}{2}\Sigma_j^k} = \cos\left(\alpha/2\right)\mathbb{I} + \iu\sin\left(\alpha/2\right) \Sigma_j^k ,
\end{equation}
whose action on the Majorana operators (\ref{majorana op}) can be obtained from the following relation
\begin{equation}
 U_{-\alpha}^{(k)}\, \Sigma_j^{b}\, U_{\alpha}^{(k)} 
 = \big[ \delta_{k,b} + (1-\delta_{k,b}\cos\alpha) \big]\Sigma_j^b +  (\sin\alpha) \varepsilon_{kb\ell} \,\Sigma_j^\ell ,
\end{equation}
where $\varepsilon_{kb\ell} $ is the totally antisymmetric tensor such that $\varepsilon_{xyz} =1$.

In this manuscript we are interested in a subsystem $A= A_1 \cup A_2$ made by two disjoint blocks of lattice sites.
Denoting by $B= B_1 \cup B_2$ the complementary set of sites, which is also made by two disjoint blocks, 
the reduced density matrix is $\rho_A =\Tr_B \ket{\Psi}\bra{\Psi}$.
This operator is Gaussian and it can be written  as  \cite{vlrk-03} 
\begin{equation} 
\label{eq:rhoA}
 \rho_A = \frac{1}{2^{\ell_1+\ell_2}}  \sum  w_{12}  \, O_1 O_2,
 \qquad
 w_{12} = \braket{O_2^\dagger O_1^\dagger},
\end{equation}
where $O_k$ (with $k\in \{1,2\}$) is a generic product of Majorana operators in $A_k$, namely $O_k = \prod_{j\in A_k}(a^x_{j})^{\mu^x_{[j]}}(a^y_{j})^{\mu^y_{[j]}}$ with $\mu_{[j]}^\alpha\in \{0,1\}$.
The sum in (\ref{eq:rhoA}) is performed over all possible combinations of $\mu_{[j]}^\alpha$.

Let us consider the operator $O_2$ and introduce $\mu_2=\sum_{j\in A_2}(\mu_{[j]}^x+\mu_{[j]}^y)$ the total number of Majorana operators in $O_2$ and $\mu_2^y=\sum_{j\in A_2}\mu_{[j]}^y$ the number of $a_r^y$'s in $O_2$.
The transpose of $O_2$ (which obviously coincides with the partial transpose with respect to $A_2$ in this case) is given by
\begin{equation}
\label{O2 transp v1}
 O_2^T 
 = (-1)^{\tau(\mu_2)} (-1)^{\mu_2^y} \,O_2 ,
\end{equation}
where
\begin{equation}
\label{tau mu2}
 \tau(\mu_2) = 
 \begin{cases}
  0 & (\mu_2 \textrm{ mod } 4) \in \{0,1\}, \\
  1 & (\mu_2 \textrm{ mod } 4) \in \{2,3\} .
 \end{cases}
\end{equation}
The factor $(-1)^{\tau(\mu_2)}$ in (\ref{O2 transp v1}) originates from a rearrangement of the $a^{x,y}$'s operators after the transposition,
while the factor $(-1)^{\mu_2^y}$ comes from the fact that $(a_r^y)^T = -a_r^y$ and $(a_r^x)^T = a_r^x$ for the Majorana operators occurring in $O_2$.
This extra factor can be removed by a unitary transformation.
Another transposition can be naturally defined, namely
\begin{equation}
\label{hat transposition def}
 O_2^{\widehat T} = U_{-\pi}^{(x)}\, O_2^T\, U_{\pi}^{(x)} = (-1)^{\tau(\mu_2)} O_2,
\end{equation}
where the unitary $U_{\pi}^{(x)}$ is now a product of terms like (\ref{Uk def}) over all the sites and it changes the sign of the $a_r^y$'s leaving the $a_r^x$'s untouched.
This is the definition introduced in \cite{ez-15} and we will adopt this convention throughout this manuscript.
Thus, let us drop the hat in (\ref{hat transposition def}) and denote it simply by $O_2^T$.

Given a block $C$ of contiguous sites, an important ingredient in our analysis is the following string of Majorana operators
\begin{equation}
\label{P_C def}
 P_{C} = \prod_{j\in C} \iu \, a^x_j a^y_j ,
\end{equation}
which satisfies $P_C^{-1} =P_C$.

The partial transpose of $\rho_A$ with respect to  $A_2$ can be written as the following sum of two Gaussian operators \cite{ez-15}
\begin{equation} 
\label{eq:ez}
 \rho_A^{T_2} 
 = \frac{1-\iu}{2}\,\tilde\rho_A + \frac{1+\iu}{2}\,P_{A_2} \tilde\rho_A P_{A_2}
 = \frac{1}{\sqrt 2}\, \big( e^{-\iu\frac{\pi}{4}}\tilde{\rho}_A + e^{\iu\frac{\pi}{4}} P_{A_2}  \tilde{\rho}_A P_{A_2} \big),
\end{equation}
where the construction of $\tilde\rho_A $ has been reviewed in App.~\ref{app: rdm} and  $P_{A_2} $ is the string of Majorana operators (\ref{P_C def}) along $A_2$. 

The computation of $\Tr (\rho_A^{T_2})^n$ through (\ref{eq:ez}) provides an expression containing $2^n$ terms given by 
all the combinations of $\tilde{\rho}_0\equiv\tilde\rho_A$ and $\tilde{\rho}_1\equiv P_{A_2}\tilde\rho_A P_{A_2}$, which can 
be written as 
\be
\label{pt A2 sum}
\textrm{Tr}\big(\rho_A^{T_2}\big)^n
=
\sum_{p_1,p_2,\dots p_n=0,1}\, 
  \frac{e^{\iu \frac{\pi}{4} \sum_{i}^n p_i } \, e^{-\iu \frac{\pi}{4} \left( n-\sum_{i}^n p_i \right)}}{2^{n/2}}
  \;
 \Tr\bigg[ \,\prod_{k=1}^n\tilde{\rho}_{p_k} \bigg]. 
\ee
This formula can be further simplified noticing that the various terms in the sum are invariant under the exchange $p_i\to 1-p_i$.
Using this and reorganising the terms in the sum, we can write 
\be
\label{pt A2 sum v1}
\textrm{Tr}(\rho_A^{T_2})^n
=
\frac{1}{2^{n-1}}
\sum_{\boldsymbol{p}}\, 
  2^{n/2} \cos\bigg[  \frac{\pi}{4} \bigg( 2\sum_{i=1}^{n-1} p_i-n \bigg) \bigg]  \,
 \Tr\bigg[ \tilde{\rho}_{0} \prod_{k=1}^{n-1}\tilde{\rho}_{p_k} \bigg] ,
\ee
where the vector $\boldsymbol{p}$ has $n-1$ components equal to $0$ or $1$ and therefore the sum contains $2^{n-1}$ terms.

\subsection{Fermionic coherent states for a single site}
\label{subsec coherent state site}

In this subsection, we briefly review the features of the fermionic coherent states \cite{kleinert} which are needed to 
build the path integral of $\rho_A$ and $\tilde\rho_A$.
Here we focus on a single site (indeed, the site index will be dropped in this subsection) and in the next subsection the natural extension to many sites will be considered.

The coherent states for fermions are defined through the Grassmann anticommuting variables.
If $\theta_1$ and $\theta_2$ are real Grassmann variables, we have that $\theta_i^2=0$ for $i\in \{1,2\}$ and $\theta_1\theta_2 = -\theta_2\theta_1$.
Since $\theta^2=0$,  a function $f(\theta)$ of the real Grassman variable can be written as $ f(\theta) = f_0 + f_1\theta$.
Given two real Grassmann variables one can build a complex Grassmann variable $\zeta$ as follows
\begin{equation}
 \zeta = \frac{1}{\sqrt 2}\big( \theta_1+\iu\,\theta_2 \big),
 \qquad 
 \conj\zeta = \frac{1}{\sqrt 2}\big( \theta_1-\iu\,\theta_2 \big).
\end{equation}
The integration over a complex Grassmann variable acts as a derivation; indeed
\begin{equation}
 \int\df\zeta = 0,
 \quad \int\df\zeta \,\zeta = 0,
 \quad 
 \int\df\zeta \,\conj\zeta = 0,
 \quad   \int\df\zeta \,\zeta\,\conj\zeta = 1.
\end{equation}

The coherent states are defined as follows
\begin{equation}
\label{coherent states}
 \ket{\zeta} = \ket{0} - \zeta\ket{1} ,
 \qquad \bra{\zeta} = \bra{0} + \conj\zeta\bra{1} .
\end{equation}
Since $\zeta$ commutes with $\ket{0}$ and anticommutes with $c$, $c^\dagger$ and $\ket{1}$, it is straightforward to check that $c\ket{\zeta} = \zeta\ket{\zeta} $ and $\bra{\zeta}c^\dag = \bra{\zeta} \conj{\zeta}$.
Notice that the coherent states do not provide an orthonormal basis. 
A completeness relation and a formula for the trace of an operator $O$ read respectively
\begin{equation} 
\label{eq:identity and trace}
 \mathbb{I} = \intf{\zeta}\,e^{-\conj\zeta\zeta}\ket{\zeta}\bra{\zeta},
 \qquad 
 \textrm{Tr}\, O = \intf{\zeta}\,e^{-\conj\zeta\zeta}\bra{-\zeta} O\ket{\zeta} .
\end{equation}

Given the above rules, the matrix elements of the identity and of the operators in (\ref{majorana op}) on the 
coherent states (\ref{coherent states}) can be computed, finding that
\bea
\label{id cs}
  & &\braket{\zeta|\eta} = 1+\conj\zeta\eta = \braket{\conj\eta|-\conj\zeta} ,\\
\label{1 cs}
  & & \braket{\zeta| a^x |\eta} = \conj\zeta+\eta = \braket{\conj\eta| a^x |\conj\zeta}, \\
  \label{2 cs}
  & & \braket{\zeta| a^y |\eta} = -\iu(\conj\zeta-\eta) = -\braket{\conj\eta| a^y |\conj\zeta} ,\\
    \label{3 cs}
    & & \braket{\zeta|\iu a^x a^y|\eta} = 1-\conj\zeta\eta = \braket{\conj\eta|\iu a^x a^y|-\conj\zeta} ,
  \label{last}  
\eea
where the second rewriting will be useful in the following subsection. 
Since $\iu\,a^x a^y\ket{\zeta} = \ket{-\zeta}$, we can bring (\ref{1 cs}) and (\ref{2 cs}) in the same form of (\ref{id cs}) and (\ref{3 cs}):
\bea
\label{4 cs}
  & & \braket{\zeta| a^x |\eta} 
  = \braket{\conj\eta| \iu a^y |-\conj\zeta}  = -\iu\braket{\conj\eta| \,U_{-\pi/2}^{(z)}\, a^x \,U_{\pi/2}^{(z)}\, |-\conj\zeta}, \\
  \label{5 cs}
  & & \braket{\zeta| a^y |\eta} 
  = \braket{\conj\eta| \iu a^x |-\conj\zeta}  = \phantom{-} \iu\braket{\conj\eta| \,U_{-\pi/2}^{(z)}\, a^y \,U_{\pi/2}^{(z)}\, |-\conj\zeta} .
\eea
Notice that the insertion of $U_{\pi/2}^{(z)}$ and its hermitian conjugate in (\ref{id cs}) and (\ref{3 cs}) has no effect.

\subsection{Partial transpose of the reduced density matrix}
\label{subsec pt coherent}

The coherent state $\ket{\zeta(x)}$ for a lattice is the tensor product of single site coherent states, with $x$ runnig along the whole system or the corresponding subsystem.
In the following we consider a lattice system but the final formulas can be extended to a continuous spatial dimension in a straightforward way by interpreting the discrete sums as integrals and integrations over a discrete set of variables as path integrals.

\begin{figure}
\vspace{.6cm}
\includegraphics[width=1.\textwidth]{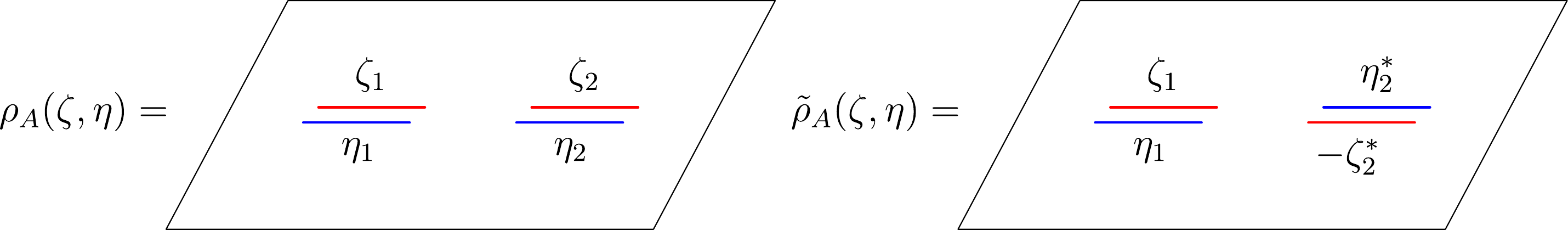}
\caption{Path integral representation of $\rho_A$ and $\tilde \rho_A$ in the coherent state basis.
} 
\label{fig:shee}
\end{figure}

The density matrix of the whole system in the ground state is $\rho = \ket{\Psi}\bra{\Psi}$ and its matrix element between two generic coherent states reads
\begin{equation}
\label{rho mat element}
 \rho(\zeta(x),\eta(x)) \,=\,  e^{-\conj\zeta \eta} \braket{\zeta(x)|\Psi}\braket{\Psi|\eta(x)},
\end{equation}
where $\conj\zeta \eta= \sum_x \conj\zeta(x) \, \eta(x)$, with $x$ labelling the whole system and $e^{-\conj\zeta \eta} $ is the normalization factor (see (\ref{id cs})).
To obtain the reduced density matrix in $A$, one first separates the degrees of freedom in $A$ and the ones in $B$ and then traces over the latter ones. 
Denoting by $\ket{\zeta_A(x_A)}$ and $\ket{\zeta_B(x_B)}$ the coherent states on $A$ and $B$ respectively, we have that $ \ket{\zeta(x)} =  \ket{\zeta_A(x_A)}\otimes \ket{\zeta_B(x_B)} $.
Adopting the notation $ \ket{\zeta(x)}  = \ket{\zeta_A(x_A),\,\zeta_B(x_B)} $, the matrix element of $\rho_A$ is given by
\begin{equation}
 \rho_A(\zeta_A(x_A),\eta_A(x_A)) \,=\, 
 e^{-\conj\zeta_A \eta_A} 
 \int D\chi_B^\ast \, D\chi_B
 \, e^{-\conj\chi_B \,\chi_B} \braket{\zeta_A,\,-\chi_B|\Psi}  \braket{\Psi|\eta_A,\,\chi_B} ,
\end{equation}
where $D \chi_B^\ast \, D\chi_B =   \prod_{x_B} {\rm d}\chi_B^\ast(x_B) \, {\rm d}\chi_B(x_B)$ and the minus sign comes from the trace over $B$, according to (\ref{eq:identity and trace}). 
In the continuum limit, the braket $\braket{\Psi|\eta_A,\,\chi_B}$ is the fermionic path integral on the upper half plane where the boundary conditions $\eta_A(x_A)$ and $\chi_B(x_B)$ are imposed in $A$ and $B$ respectively, just above the real axis.
Analogously, $\braket{\zeta_A,\,-\chi_B|\Psi}$ is the path integral on the lower half plane.
The trace over $B$ is performed by setting the fields along $B$ equal (but with opposite sign) and summing over all the configurations.
The resulting path integral is over the whole plane with two open slits along $A_1$ and $A_2$, where the boundary conditions $\eta_A$ and $\zeta_A$ are imposed respectively along the lower and the upper edge of $A$ (left panel of Fig.~\ref{fig:shee}).

Let us consider the partial transpose $\rho_A^{T_2}$  with respect to $A_2$.
In App.~\ref{app: rdm} the lattice results of \cite{ez-15} that we need in our analysis are briefly reviewed.
Remembering that the partial transposition acts only on operators in $A_2$, from (\ref{eq:rhoA}) we can write its matrix elements as follows
\begin{equation} 
\label{eq:rho_sum}
 \braket{\zeta(x)|\rho_A^{T_2}|\eta(x)} 
 = \frac{1}{2^{\ell_1+\ell_2}} 
 \sum w_{12}\,  \braket{\zeta_1(x_1)|O_1|\eta_1(x_1)}\braket{\zeta_2(x_2)|O^T_2|\eta_2(x_2)} ,
\end{equation}
where $x_j \in A_j$, with $j\in \{1,2\}$.

Focussing on the term corresponding to  $A_2$ in (\ref{eq:rho_sum}), from (\ref{hat transposition def}), (\ref{id cs})-(\ref{5 cs}) one finds
\begin{equation}
\label{O2T}
  \braket{\zeta_2(x_2)|O^T_2|\eta_2(x_2)} 
  = (-1)^{\tau(\mu_2)} \, \iu^{\mu_2^y-\mu_2^x}\braket{\conj\eta_2(x_2)|\,U_{-\pi/2}^{(z)}\,O_2\,U_{\pi/2}^{(z)}\,|-\conj\zeta_2(x_2)}  ,
\end{equation}
where the unitary map $U_{-\pi/2}^{(z)}$ acts on all sites.
When the number $\mu_2$ of Majorana operators in $A_2$ is even, from (\ref{tau mu2}) we have that $(-1)^{\tau(\mu_2)} = \iu^{\mu_2}$ and therefore
\bea
\fl
 \braket{\zeta_2(x_2)|(O^T_2)_{\text{even}}|\eta_2(x_2)} 
 &=& (-1)^{\mu_2^y} \braket{\conj\eta_2(x_2)|\,U_{-\pi/2}^{(z)}\,(O_2)_{\text{even}}\,U_{\pi/2}^{(z)}\,|-\conj\zeta_2(x_2)} 
 \\
 \label{O2trans even}
 & = & 
 \braket{\conj\eta_2(x_2)|\,U_{-\pi}^{(y)}\, U_{-\pi/2}^{(z)}\,(O_2)_{\text{even}}\,U_{\pi/2}^{(z)}\,U_{\pi}^{(y)}\,|-\conj\zeta_2(x_2)},
\eea
where in (\ref{O2trans even}) the factor $(-1)^{\mu_2^y}$ has been removed through a second unitary transformation which sends $a^x_j\rightarrow -a^x_j$ leaving the $a^y_j$'s unchanged (we recall that $U_{-\pi/2}^{(z)}$ exchanges the $a^x_j$'s with the $a^y_j$'s).
The expression (\ref{O2trans even}) suggests us to introduce the following unitary operator acting on $A_2$ 
\begin{equation}
 V_2 \equiv U_{-\pi}^{(y)} \, U_{-\pi/2}^{(z)}
 = \prod_{j\in A_2}\exp\left(-\,\iu\,\frac{\pi}{2}\,\frac{a_j^x-a_j^y}{\sqrt 2}\right) 
 = \prod_{j\in A_2} \exp\left[-\, \iu\,\frac{\pi}{2}\left(e^{\iu\frac{\pi}{4}}c_j^\dag+e^{-\iu\frac{\pi}{4}}c_j\right)\right] ,
\end{equation}
whose net effect is to send $a^x_j\rightarrow -a^y_j$ and $a^y_j\rightarrow -a^x_j$, for $j\in A_2$.

In a similar way, we can treat the case of odd $\mu_2$, for which $(-1)^{\tau(\mu_2)} = \iu^{\mu_2-1}$ (see (\ref{tau mu2})).
Again, from (\ref{hat transposition def}), (\ref{id cs})-(\ref{5 cs}) one gets
\bea
\fl
 \braket{\zeta_2(x_2)|\big(O^T_2\big)_{\text{odd}}|\eta_2(x_2)} 
 &=& -\iu \, (-1)^{\mu_2^y} \braket{\conj\eta_2(x_2)|\,U_{-\pi/2}^{(z)}\,(O_2)_{\text{odd}}\,U_{\pi/2}^{(z)}\,|-\conj\zeta_2(x_2)} 
 \\
 &=& -\iu \braket{\conj\eta_2(x_2)|\,V_2\,(O_2)_{\text{odd}}\,V_2^\dag\,|-\conj\zeta_2(x_2)} .
\eea
Introducing the operator $\widetilde O_2$ through its matrix elements as follows
\begin{equation}
 \braket{\zeta_2(x_2)|\widetilde O_2|\eta_2(x_2)} = \braket{\conj\eta_2(x_2)|\,V_2\,O_2\,V_2^\dag\,|-\conj\zeta_2(x_2)} ,
\end{equation}
the expression (\ref{eq:rho_sum}) can be written as follows
\bea
\label{rhoT2 sandwich}
& &\hspace{-1.5cm}
 \braket{\zeta(x)|\rho_A^{T_2}|\eta(x)}
 \,=\, \frac{1}{2^{\ell_1+\ell_2}}  \bigg( \sum_{\text{even}}w_{12}\,\braket{\zeta_1(x_1)|O_1|\eta_1(x_1)}\braket{\zeta_2(x_2)|\widetilde O_2|\eta_2(x_2)} \\
& &\hspace{3.5cm} 
- \iu \sum_{\text{odd}}w_{12}\,\braket{\zeta_1(x_1)|O_1|\eta_1(x_1)}\braket{\zeta_2(x_2)|\widetilde O_2|\eta_2(x_2)} \bigg),
\nonumber
\eea
where in the first (second) sum the terms have an even (odd) number of fermionic operators in $A_2$ (see App.~\ref{app: rdm}).
In (\ref{rhoT2 sandwich}) the structure $ \rho_A^{T_2} = \tilde\rho_{\rm even} - \iu \, \tilde\rho_{\rm odd}$ (see (\ref{eq:rhoT even odd})) can be recognised and this observation leads us to identify the matrix element of $\tilde\rho_A$ on the coherent states
\begin{equation} 
\label{eq:rhotilde matrix element}
 \braket{\zeta(x)|\tilde\rho_A|\eta(x)} = \braket{\zeta_1(x_1),\,\conj\eta_2(x_2)|\,V_2\,\rho_A \,V_2^\dag\,|\eta_1(x_1),\,-\conj\zeta_2(x_2)} ,
\end{equation}
and analogously 
\begin{equation} 
\label{eq:rhotilde matrix element2}
 \braket{\zeta(x)|P_{A_2} \tilde\rho_A P_{A_2}|\eta(x)} = 
 \braket{\zeta_1(x_1),\,-\conj\eta_2(x_2)|\,V_2\,\rho_A \,V_2^\dag\,|\eta_1(x_1),\,\conj\zeta_2(x_2)} ,
\end{equation}
where from (\ref{last}) we can read that the action of $P_{A_2}$ is to change the sign of $\zeta_2$. 
A graphical representation of this path integral representation for $\tilde\rho_A$ is given in the right panel of Fig.~\ref{fig:shee}.
Hence, the final expression for the the partial transpose in the coherent state basis can be written exactly like the lattice counterpart i.e.
\begin{equation}
\rho_A^{T_2}(\zeta,\eta)
= \frac{1}{\sqrt 2}\, \Big[ e^{-\iu\frac{\pi}{4}}\tilde{\rho}_A(\zeta_1,\zeta_2; \eta_1,\eta_2)
 + e^{\iu\frac{\pi}{4}}   \tilde{\rho}_A (\zeta_1,-\zeta_2; \eta_1,-\eta_2) \Big],
\label{pathin}
\end{equation}
where the notation is such that $\eta_i,\zeta_i\in A_i$.
This explicit form of the partial transpose in the coherent state basis is the final and main result of this section. 

In App.~\ref{app check n=2} we employ the formalism of this section to check the identity $\Tr ( \rho_A^{T_2} )^2 = \Tr \rho_A^2$ (which holds for any quantum system \cite{cct-neg-letter, cct-neg-long}) for the free fermion.

\section{Traces of the partial transpose for the free fermion}
\label{sec:FF}

In this section we consider the moments $\textrm{Tr}(\rho_A^{T_2})^n$ for the free fermion.
After summarising some needed CFT results for the moments of $\rho_A$ (Sec.\,\ref{mom1}), 
using the path integral approach of the previous section, 
we derive the analytic formula for $\textrm{Tr}(\rho_A^{T_2})^n$ given by (\ref{eq:partial transpose free fermion})  
which is the main result of this manuscript.
In Sec.~\ref{subsec:boson eta1} we show that the moments for the free fermion are equal to the ones for the compact boson 
at the self-dual radius. 
Finally, in Sec.~\ref{subsec:FF numerics} we give some numerical checks of our results. 

\subsection{Some CFT results for $\Tr \rho_A^n$}
\label{mom1}

For the case of a single interval of length $\ell $ embedded in a CFT on the infinite line, the moments of the reduced density matrix 
can be written as \cite{Holzhey,cc-04,cc-rev}
\be
\Tr\rho_A^{n}= c_n \left(\frac{\ell}{a}\right)^{-c(n-1/n)/6},
\label{Renyi:asymp}
\ee
where $c$ is the central charge and $a$ the inverse of an ultraviolet cutoff (e.g.\ the lattice spacing).
The prefactors $c_n$ are non universal constants (that however satisfy universal relations \cite{fcm-10}). 
The simple universal dependence on the central charge can be understood because  $\Tr\rho_A^{n}$ is the 
partition function on a surface of genus zero that can be mapped to the complex plane \cite{cc-04}.
Eq.~\eqref{Renyi:asymp} can be interpreted as the two-point function of some twist operators acting at the endpoints of the interval
$u$ and $v$ \cite{cc-04,ccd-09}, i.e.\ $\Tr\rho_A^{n}= \langle {\cal T}_n(u)\bar {\cal T}_n(v)\rangle$. 
The twist fields ${\cal T}_n$ behave like primary operators with scaling dimension
\be
\Delta_n= \frac{c}{12}\left(n-\frac1n\right) .
\ee
The knowledge of the moments $\Tr \rho_A^n$ give access to the full spectrum of the reduced density matrix \cite{cl-08}.
While $c_n$ is not universal, its value for the tight-binding model at half-filling is known  exactly and it is given by \cite{jk-04}
\begin{equation}
\label{c_n free fermion}
 c_n = 2^{-\frac{1}{6}\left(n-\frac{1}{n}\right)}
 \exp\bigg\{ \,\iu\, n\int_{-\infty}^\infty \mathrm{d}z\, 
 \log\bigg(\frac{\Gamma\left(\frac{1}{2}+\iu z\right)}{\Gamma\left(\frac{1}{2}-\iu z\right)}\bigg) 
 \big[\tanh\left(\pi z\right) - \tanh\left(\pi n z\right) \big]\bigg\}.
\end{equation}

For the case of two disjoint intervals $A=A_1\cup A_2=[u_1,v_1]\cup [u_2,v_2]$, 
by global conformal invariance, in the thermodynamic limit,  $\Tr \rho_A^n$ can be written as
(dropping hereafter the dependence on the UV cutoff $a$)
\be\fl
\Tr \rho_A^n
=c_n^2 \left(\frac{|u_1-u_2||v_1-v_2|}{|u_1-v_1||u_2-v_2||u_1-v_2||u_2-v_1|} 
\right)^{2\Delta_n} \mathcal{F}_n(x),
\label{Fn}
\ee 
where $x$ is the four-point ratio (for real $u_j$ and $v_j$, $x$ is real) 
\be
\label{cross ratio def}
x =\frac{(u_1 -v_1)(u_2-v_2)}{(u_1 - u_2)(v_1 - v_2)} \,\in\, (0,1).
\ee
The function $\mathcal{F}_n(x)$ is a universal function (after being normalized such that $\mathcal{F}_n(0)=1$) that encodes all the information about 
the operator spectrum of the CFT 
while $c_n$ is the same non-universal constant  appearing in  (\ref{Renyi:asymp}).
The function $\mathcal{F}_n$ has been studied in several 
papers \cite{cct-09,cct-11,cg-08,ch-05,fps-08,c-10,headrick,hlr-13,ctt-14,atc-10,atc-11,rg-12,f-12,cz-13} 
(see \cite{RT, headrick, hol} for the holographic viewpoint and \cite{cft-high-dims} for higher dimensional conformal field theories).
In the case of two disjoint intervals, $\Tr \rho_A^n$ is the partition function on a surface of genus $n-1$ 
which cannot be mapped to the complex plane. 
This surface is usually called ${\cal R}_n$.

One of the most important examples of exactly known $\mathcal{F}_n(x)$  is the 
free  boson compactified on a circle of radius $r_{\rm circle}$.  
In this case, the function $\mathcal{F}_n(x)$ (parametrized in terms of $\eta= 2r_{\rm circle}^2$) is \cite{cct-09}
\begin{equation}
\mathcal{F}_n(x)=
\frac{\Theta\big({\bf 0}|\eta\tau \big)\,\Theta\big({\bf 0}|\tau/\eta\big)}{
[\Theta\big({\bf 0}|\tau\big)]^2},
\label{Fnv}
\end{equation}
where $\tau$ is an $(n-1)\times(n-1)$ matrix (called period matrix) with elements \cite{cct-09}
\begin{equation} 
\label{eq:tau}
 \tau_{i,j} \,=\, \iu\,  \frac{2}{n} \sum_{k=1}^{n-1} \sin(\pi k/n) \, \frac{{}_2F_1(k/n,1-k/n;1;1-x)}{{}_2F_1(k/n,1-k/n;1;x)} \, \cos[2\pi (k/n)(i-j)].
\end{equation}
We remark that, since $x\in (0,1)$, the period matrix $ \tau(x)$ is purely imaginary. 
$\Theta$ is the Riemann theta function \cite{theta books,Fay book}
\begin{equation}
\label{theta Riemann def 0}
\Theta(\boldsymbol{z}|M)\,\equiv\,
\sum _{\boldsymbol{m}\,\in\,\mathbb{Z}^{n-1}}
e^{
\,{\rm i} \pi \, \boldsymbol{m}^{\rm t} \cdot M \cdot \boldsymbol{m}
+2\pi {\rm i} \, \boldsymbol{m}^{\rm t}\cdot \boldsymbol{z} 
},
\end{equation}
which is a function of the $(n-1)$ dimensional complex vector $\boldsymbol{z}$  and of 
the $(n-1) \times (n-1)$ matrix  $M$ which must be symmetric and with positive imaginary part.

For the critical Ising model, the scaling function $\mathcal{F}_n (x)$ is also known \cite{cct-11} 
\begin{equation}
\mathcal{F}_n (x)= 
\frac{1}{2^{n-1}\, \Theta({\bf 0}|\tau)} \sum_{\bm{\varepsilon,\delta}}
\bigg| \Theta\bigg[\hspace{-.1cm}
\begin{array}{c} \bm{\varepsilon} \\ \bm{\delta}  \end{array}
\hspace{-.1cm}\bigg] ({\bf 0}|\tau)\bigg|,
\label{Fn ising}
\end{equation}
where the period matrix  $\tau$ is the same as in Eq.~\eqref{eq:tau}.
In this case $\Theta$ is the Riemann theta function with characteristic defined as \cite{theta books, Fay book}
\be\fl
\label{riemann theta def}
\Theta[\boldsymbol{e}](\boldsymbol{z}|M) \,
\equiv
\sum _{\boldsymbol{m}\,\in\,\mathbb{Z}^{n-1}}
e^{
\,{\rm i} \pi
(\boldsymbol{m}+\boldsymbol{\varepsilon})^{\rm t}
\cdot M \cdot(\boldsymbol{m}+\boldsymbol{\varepsilon})
+2\pi {\rm i} \,
(\boldsymbol{m}+\boldsymbol{\varepsilon})^{\rm t}\cdot (\boldsymbol{z} + \boldsymbol{\delta})
},
\qquad
\boldsymbol{e}  
\equiv 
\bigg( \hspace{-.1cm}
\begin{array}{c}
 \boldsymbol{\varepsilon}\\
 \boldsymbol{\delta}
 \end{array}\hspace{-.1cm}\bigg),
\ee
where $\boldsymbol{z}$ and $M$ are analogous to the ones in  (\ref{theta Riemann def 0}), and 
$\bm{\varepsilon}, \bm{\delta} $ are vector with entries  $0$ and $1/2$.
The sum in $(\bm{\varepsilon,\delta})$ in (\ref{Fn ising}) is intended over all the $2^{n-1}$ vectors 
${\bm \varepsilon}$ and ${\bm \delta}$ with these entries. 
The parity of (\ref{riemann theta def}) as function of $\boldsymbol{z}$ is given by the parity of the characteristic, 
which is the parity of the integer number $4\boldsymbol\varepsilon\cdot\boldsymbol\delta$.
There are $2^{2(n-1)}$ characteristics: $2^{n-2}(2^{n-1}+1)$ are even and $2^{n-2}(2^{n-1}-1)$ are odd. 
In our following  analysis only the trivial vector $\boldsymbol{z}=\boldsymbol{0}$ occurs and therefore we will adopt the shortcut notation:
$\Theta[{\bf e}](M)  \equiv \Theta[\boldsymbol{e}](\boldsymbol{0}|M) $ and $\Theta(M)  \equiv \Theta(\boldsymbol{0}|M) $ when the characteristic is vanishing. 

\begin{figure}
\vspace{.6cm}
\includegraphics[width=1.\textwidth]{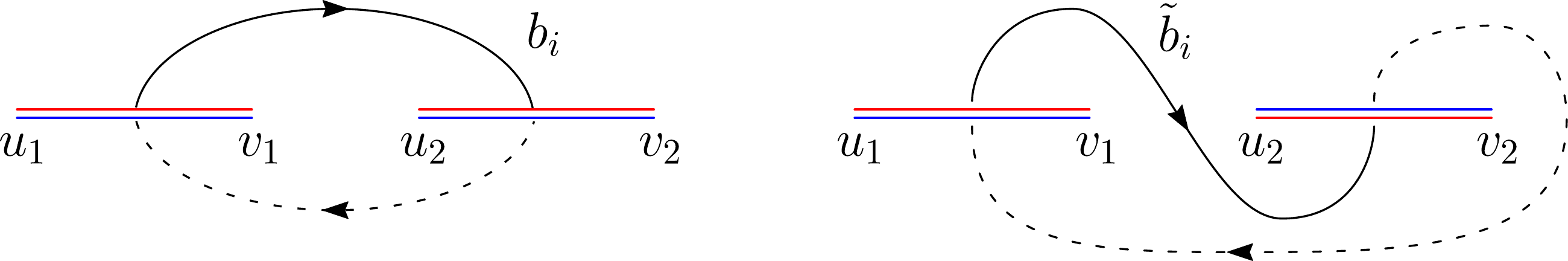}
\caption{Cycles of type $b$ for $\mathcal{R}_n$ (left) and $\widetilde{\mathcal{R}}_n$ (right).
The solid line represents the part of the cycle belonging to the $i$-th sheet, while the dashed curve is the remaining part of the cycle, which lies in the $(i+1)$-th sheet. As for the cycles of type $a$, which are the same for $\mathcal{R}_n$ and $\widetilde{\mathcal{R}}_n$, we refer to Fig.~8 of \cite{ctt-14}.
} 
\label{fig:cycles b}
\end{figure}

In the computation of the partition function on higher genus Riemann surfaces, one has to properly choose a canonical homology 
basis (i.e.\ a set of $2(n-1)$ closed 
oriented curves on the surface, the $a$ and $b$ cycles, which satisfy some specific intersection rules) and a set of 
$n-1$ holomorphic differentials.
By integrating such differentials along the $b$ cycles one gets the period matrix of the Riemann surface.
For a genus $g$ Riemann surface, the period matrix is a $g \times g$ complex symmetric matrix with positive definite imaginary part \cite{bosonization higher genus,dvv}.  
We refer the reader to \cite{ctt-14} for a detailed analysis about the canonical homology basis for $\mathcal{R}_n$.
In particular, the canonical homology basis $\{a_r, b_r \,; \, 1\leqslant r\leqslant n-1\}$ corresponding to \eqref{eq:tau} has been 
discussed in Sec.~4 of \cite{ctt-14} and we will adopt it throughout this manuscript.
In the left panel of Fig.~\ref{fig:cycles b} we show the  $j$-th $b$ cycle, which belongs to the $j$-th sheet and to the $(j+1)$-th sheet. 
Instead, the construction of $a_j$ (which intersects $b_j$ only once) is more involved and therefore we refer the interested reader to 
Fig.~8 of \cite{ctt-14}.

The Riemann theta function with characteristic (\ref{riemann theta def}) occurs in the computation of fermionic models on higher genus Riemann surfaces \cite{bosonization higher genus,dvv}.
The characteristic $ \boldsymbol e $ specifies the set of boundary conditions along the $a$ and $b$ cycles of the canonical homology basis and this provides the so called {\it spin structures} of the model. 
The vector $\boldsymbol{\varepsilon}$ is determined by the boundary conditions along the $a$ cycles ($\varepsilon_k = 0$ for antiperiodic b.c.\ around $a_k$ and $\varepsilon_k = 1/2$ for periodic b.c.), while $\boldsymbol{\delta}$ is provided by the boundary conditions along the $b$ cycles ($\delta_k = 0$ for antiperiodic b.c.\ around $b_k$ and $\delta_k = 1/2$ for periodic b.c.).

\subsection{Moments of the partial transpose for the free fermionic field theory}
\label{sec cont limit free}

We are finally ready to derive the moments of the partial transpose $\textrm{Tr}(\rho_A^{T_2})^n$.
The path integral for $\rho_A^{T_2}$ is given by (\ref{pathin}) which is a sum of two different operators. 
The moments are then given by the sum of $2^n$ terms that come from the expansion of the binomial. 
Actually, since there is a double degeneration of these terms, the sum is only on $2^{n-1}$ terms.
Introducing, in analogy with the lattice computation, the notation $\tilde \rho_0(\zeta,\eta)=\tilde \rho_A(\zeta,\eta)$ and 
$\tilde \rho_1(\zeta,\eta) = \tilde \rho_A (\zeta_1,-\zeta_2; \eta_1,-\eta_2)=P_{A_2}\tilde\rho_AP_{A_2}$, 
the $2^{n-1}$ terms in the sum for the moment of order $n$
can be written as 
\begin{equation} 
 \int\prod_{k=1}^n D\chi^*_kD\chi_k \,\tilde\rho_0(-\chi_n,\chi_1)\prod_{k=1}^{n-1} \tilde \rho_{p_k} (\chi_k,\chi_{k+1}),
 \label{terms}
\end{equation}
with $p_i=0,1$.
Each of these $2^{n-1}$ terms is a partition function of a free fermion on a Riemann surface of genus
$n-1$ in which antiperiodic or periodic boundary condition are imposed along the basis cycles.

At this point, before deriving the final result, we should discuss the Riemann surface $\widetilde{\mathcal{R}}_n$ on which 
these partition functions are defined. 
The surface $\widetilde{\mathcal{R}}_n$ is defined by the density matrix $\tilde \rho_A$ and  
has genus $n-1$ \cite{cct-neg-letter, cct-neg-long}, but it is different 
from the one defining $\textrm{Tr}\rho_A^n$ (denoted in the previous section as $\mathcal{R}_n$). 
Only for $n=2$ they are the same torus (their moduli are related by a modular transformation), but for $n>2$ they are different. 
The properties of this Riemann surface are discussed in details in App.~\ref{app:Rn}.
The period matrix $\tilde{\tau}(x)$ of $\widetilde{\mathcal{R}}_n$ for $x\in (0,1)$ is given by \cite{cct-neg-long}
\begin{equation} 
\label{eq:Q}
\tilde{\tau}(x) = \tau\big(x/(x-1)\big) = \mathcal{R} + \iu \,\mathcal{I} ,
\qquad \mathcal{R} = \frac{1}{2}\, \mathcal{Q},
\end{equation}
where the elements of $\tau$ have been defined in (\ref{eq:tau}) and the real and imaginary parts of $\tilde{\tau}(x)$ 
are $\mathcal{R}$ and $\mathcal{I} $ respectively. 
Here it is important to observe that $\mathcal{Q}$ is a very simple symmetric integer matrix: 
it has $2$ along the principal diagonal, $-1$ along the secondary diagonals and $0$ for the remaining elements.
In App.~\ref{app:Q} we report the detailed derivation of this result.
 As for the cycles of $\widetilde{\mathcal{R}}_n$ providing the canonical homology basis 
 $\{\tilde{a}_r, \tilde{b}_r \,; \, 1\leqslant r\leqslant n-1\}$ which gives the period matrix \eqref{eq:Q}, we find 
 that $\tilde{a}_r $ is the same as $a_r$ (we remind that $\widetilde{\mathcal{R}}_n$ and $\mathcal{R}_n$ differ only 
 for the way to join the sheets along $A_2$), while the generic cycle $\tilde{b}_r $ is obtained by deforming the cycle $b_r $
as shown in Fig.~\ref{fig:cycles b}.


An important ingredient at this point is the operator $P_{A_2}$.
For an arbitrary interval $C$ we can write
\begin{equation}
\label{P_C cont limit}
 P_{C}=  (-1)^{\int_{C}\text d x\, \bar\psi(x) \psi(x)}  \equiv (-1)^{F_C}.
\end{equation}
where $F_C$ is the fermionic number operator in the interval $C$ which was already introduced long ago \cite{ginsparg}.
This operator is located along the interval $C$ and it changes the fermionic boundary conditions (from antiperiodic to periodic or viceversa) 
on a cycle whenever it crosses the curve $C$. 
In  (\ref{pathin}) for $\rho_A^{T_2}$, we have that $P_{A_2}$ occurs both before and after $\tilde{\rho}_A$.
This corresponds to the insertion of the  operators $(-1)^{F_{A_2}}$ above and below the cut along $A_2$.

Each term (\ref{terms}) is a partition function on $\widetilde{\mathcal R}_n$ with some specific boundary conditions along the $a$ and $b$ cycles and it can be expressed in terms of Riemann theta functions.
Explicitly, we have 
\begin{equation} 
\label{eq:NegFF}
 \Tr\bigg[ \tilde{\rho}_{0} \prod_{k=1}^{n-1}\tilde{\rho}_{p_k} \bigg] 
=
c_n^2 \left( \frac{1-x}{\ell_1\ell_2} \right)^{2\Delta_n}  \,  \bigg|\frac{\Theta[ \boldsymbol e ] 
(\tilde\tau(x))}{\Theta(\tilde\tau(x))}\bigg|^2,
  \qquad\hspace{.5cm}
   \boldsymbol e  = \bigg(\hspace{-.1cm}
\begin{array}{c}
 \boldsymbol{0}\\
 \boldsymbol{\delta}
 \end{array} \hspace{-.1cm} \bigg),
\end{equation}
where $\boldsymbol{0}$ is the vector made by $n-1$ zeros.
In this formula we have still to fix the vector $\boldsymbol \delta$ in terms of ${\boldsymbol p}$, which is done as follows.
Eq.~\eqref{terms} is evaluated on the $n$-sheeted Riemann surface $\widetilde{\mathcal{R}}_n$ where the $i$-th sheet is associated to the 
$\tilde{\rho}_{p_i}$. 
On the sheets associated to $\tilde{\rho}_1$, two operators $(-1)^{F_{A_2}}$ 
must be placed above and below $A_2$.
Then, the spin structure $\boldsymbol{e} $ can be read off by counting how many times the cycles of the basis cross the curves 
$A_2$. 
Since the cycles $\tilde{a}_i $ do not intersect $A_2$ at all, we have that $\boldsymbol{\varepsilon} = \boldsymbol{0} $,
 i.e.\ the boundary conditions for the fermion along  all the cycles $\tilde{a}_r $ are antiperiodic.
Instead, for $\tilde{b}_r $ this analysis is non trivial because it intersects $A_2$ on the $r$-th  sheet and on the $(r+1)$-th  
sheet, as one can see from the right panel of Fig.~\ref{fig:cycles b}.
If $\tilde{b}_r$ crosses these curves an even number of times, then $\delta_r=0$, otherwise $\delta_r=1/2$. 
It is not difficult to conclude that
\be
\label{corr delta-g}
 2\delta_i = (p_{i}+p_{i+1}) \textrm{ mod } 2,
\ee
whose inverse reads
\be
\label{p_i inverse}
 p_i = \bigg( \sum_{j=i}^{n-1} 2\delta_j \bigg) \textrm{ mod } 2
  \,=\, \frac{1-(-1)^{2\sum_{j=i}^{n-1} \delta_j}}{2} .
\ee
The simplest example of (\ref{eq:NegFF}) is the term $\textrm{Tr} \tilde{\rho}_0^{\,n}$ (namely $\boldsymbol p = \bf 0$). This spin structure has antiperiodic boundary conditions along all the cycles, i.e.\ $\boldsymbol\varepsilon =\boldsymbol\delta = \boldsymbol 0$.
For this term $\widetilde{\Omega}_n[ \boldsymbol e ]^2 =1$.

Thus, $ \Tr\left(\rho_A^{T_2}\right)^n $ can be written as a sum over all the allowed spin structures:
\begin{equation} 
\label{eq:partial transpose free fermion}
 \Tr\left(\rho_A^{T_2}\right)^n 
 =  \,
 c_n^2 \left( \frac{1-x}{\ell_1\ell_2} \right)^{2\Delta_n} \frac{1}{2^{n-1}} 
 \sum_{\boldsymbol\delta}  r_n(\boldsymbol\delta)\,
 \bigg|\frac{\Theta[ \boldsymbol e ] 
(\tilde\tau)}{\Theta(\tilde\tau)}\bigg|^2,
  \qquad
   \boldsymbol e  = \bigg(\hspace{-.1cm}
\begin{array}{c}
 \boldsymbol{0}\\
 \boldsymbol{\delta}
 \end{array} \hspace{-.1cm} \bigg).
\end{equation}
The coefficient $r_n(\boldsymbol \delta)$ is 
\begin{equation} 
\label{eq:r from EZ + correspondence}
 r_n(\boldsymbol \delta) 
 =\,
  2^{n/2}\cos\bigg[ \frac{\pi}{4} \bigg(1+\sum_{i=1}^{n-1}(-1)^{2\sum_{j=i}^{n-1} \delta_j}\bigg)\bigg] .
\end{equation}
It can be seen that  $r_n(\boldsymbol \delta) \in \{-2^{n/2} , 0, 2^{n/2}\}$ for even $n$ and 
$r_n(\boldsymbol \delta) \in \{-2^{(n-1)/2} , 2^{(n-1)/2}\}$ for odd $n$.

The analytic expression given by (\ref{eq:partial transpose free fermion}) and (\ref{eq:r from EZ + correspondence}) is 
the main result of this manuscript.
When the size of the intervals is very small with respect to their distance ($\ell_1, \ell_2 \ll d$. i.e.\ $x\ll 1$), it is possible 
to expand (\ref{eq:partial transpose free fermion}) in powers of $x$, as shown 
in App.~\ref{app: small x} where we find the first non trivial term of this expansion.


There is also a very interesting by-product of our analysis which is given by 
(\ref{eq:NegFF}) providing a very deep technical insight. 
Indeed Eq.~\eqref{eq:NegFF} shows also that {\it each of the $2^{n-1}$ terms in the sum over ${\boldsymbol p}$ in (\ref{pt A2 sum v1}) 
has a well defined continuum limit} which is 
the partition function of the free fermion on $\widetilde{\mathcal{R}}_n$ with a particular assignment of 
fermionic boundary conditions, i.e.\ always 
antiperiodic along all the $a$ cycles, while the b.c.\ along the $b$ cycles are specified by $\boldsymbol{\delta}$ 
(we recall, antiperiodic for $\delta_i=0$ and periodic otherwise).

\subsubsection{Dihedral symmetry.}
\label{subsec dih symm renyi free}

The Riemann surfaces $\mathcal{R}_n$ and $\widetilde{\mathcal{R}}_n$ enjoy a dihedral symmetry $\mathbb{Z}_n \times \mathbb{Z}_2$, as already noticed in \cite{hlr-13}.
The symmetry $\mathbb{Z}_n$ comes from the invariance under cyclic permutation of the $n$ sheets and the symmetry $\mathbb{Z}_2$ corresponds to take the sheets in the reversed order and to reflect all of them with respect to the real axis. 
The former symmetry comes  from the fact that $\mathcal R_n$ and $\widetilde{\mathcal{R}}_n$ are obtained through the replica construction and the latter one occurs because the endpoints of the intervals are on the real axis. 
Indeed, the complex equations (\ref{eq:Rn}) and (\ref{eq:Rtilden}), which define the Riemann surfaces $\mathcal{R}_n$ and $\widetilde{\mathcal{R}}_n$, are invariant under complex conjugation.

In \cite{hlr-13, ctt-14} the symplectic matrices which implement the dihedral symmetry of $\mathcal{R}_n$ have been written explicitly and in App.~\ref{app:dihn} this analysis has been extended to $\widetilde{\mathcal{R}}_n$ as well (the symmetry $\mathbb{Z}_2$ is different in the two cases).
These transformations act on the period matrix and reshuffle the characteristics, but the functions  and 
$\widetilde{\Omega}_n[ \boldsymbol e ]  $  in (\ref{tilde Jn Omega def}) remain invariant. 
Moreover, both the transformations associated to the dihedral symmetry leave the coefficient $r_n(\boldsymbol\delta)$  in \eqref{eq:r from EZ + correspondence} invariant.
Thus, the terms in the sum (\ref{eq:partial transpose free fermion}) whose characteristics are related by one of these modular transformations are equal and the sum can be written in a simpler form by choosing a representative term for each equivalence class, whose coefficient is given by (\ref{eq:r from EZ + correspondence}) multiplied by the number of terms of the  equivalence class.

Exploiting these symmetries, one can write the explicit expressions given in Sec.~6.3 of \cite{cct-15} for $2\leqslant n \leqslant 5$.
Beside the goal of having more compact analytic expressions, the dihedral symmetry is very helpful also from the numerical point of view 
because it allows to reduce the exponentially large (in $n$) number of terms in (\ref{eq:partial transpose free fermion}).

Looking at Eq.~\eqref{pt A2 sum} on the lattice, the $\mathbb{Z}_n$ symmetry corresponds to the cyclic permutation of the $n$ 
factors within each trace.
Instead, the $\mathbb{Z}_2$ symmetry comes from the fact that $\tilde{\rho}_0$ and $\tilde{\rho}_1$  are not separately hermitian but the hermitian conjugation exchange them, so that $\rho_A^{T_2}$ is hermitian.
However, as already noticed, such exchange leaves any term of the sum unchanged.

\subsection{Self-dual boson}
\label{subsec:boson eta1}

In this subsection we show that the expression (\ref{eq:partial transpose free fermion}) for $ \Tr \left( \rho_A^{T_2}\right)^n$ of the Dirac free fermion is equal to the one for the compactified boson at its self-dual radius.

The analytic formula for $ \Tr \left( \rho_A^{T_2}\right)^n$ of the compactified boson for a generic value of the compactification radius has been derived in \cite{cct-neg-long} by studying the partition function of the model on the Riemann surface $\widetilde{\mathcal R}_n$.
At the self-dual radius, it becomes (see Eq.~(146) of \cite{cct-neg-long} for $\eta=1$)
\begin{equation} 
\label{eq:RenyiNeg boson eta1}
 \Tr \left( \rho_A^{T_2}\right)^n 
 = c_n^2 \left( \frac{1-x}{\ell_1\ell_2} \right)^{2\Delta_n}
 \frac{\Theta(T)}{|\Theta(\tilde\tau)|^2},
 \qquad\quad
 T = \bigg(\hspace{-.05cm}
 \begin{array}{cc}
\iu \,\mathcal{I} & \mathcal{R} \\
\mathcal{R} & \iu \,\mathcal{I} 
\end{array}\hspace{-.05cm}\bigg),
\end{equation}
where the matrices occurring in this expression have been defined in (\ref{eq:Q}).
The Riemann theta function $\Theta(T)$ in the numerator can be written as follows
\begin{equation}
\label{eq:RenyiNeg boson eta1 step}
\Theta
\bigg(\hspace{-.05cm}
 \begin{array}{cc}
\iu \,\mathcal{I} & \mathcal{R} \\
\mathcal{R} & \iu \,\mathcal{I} 
\end{array}\hspace{-.05cm}\bigg)
=
\sum_{\boldsymbol{\varepsilon}} 
\bigg|\, \Theta\bigg[\hspace{-.1cm} \begin{array}{c}
\boldsymbol{\varepsilon} \\ \boldsymbol{0}
\end{array}\hspace{-.1cm} \bigg]
(2\tilde{\tau})\,\bigg|^2
=
\sum_{\boldsymbol{\varepsilon}} 
e^{2\pi \iu\,\boldsymbol{\varepsilon} \cdot \mathcal{Q} \cdot \boldsymbol{\varepsilon}}\;
\Theta\bigg[\hspace{-.1cm} \begin{array}{c}
\boldsymbol{\varepsilon} \\ \boldsymbol{0}
\end{array}\hspace{-.1cm} \bigg]
(2\tilde{\tau})^2,
\end{equation}
where in the first step we have used (3.6b) of \cite{dvv} and in the second one $-2\tilde{\tau}^\ast = 2\tilde{\tau} -2 \mathcal{Q}$.
Then, by specialising the addition formula reported in \cite{Fay book} (pag.\,4) to our case, we find
\begin{equation}
\label{theta final ris}
\Theta
\bigg(\hspace{-.05cm}
 \begin{array}{cc}
\iu \,\mathcal{I} & \mathcal{R} \\
\mathcal{R} & \iu \,\mathcal{I} 
\end{array}\hspace{-.05cm}\bigg)
=
\frac{1}{2^{n-1}}
\sum_{\boldsymbol{\varepsilon}, \boldsymbol{\delta}} 
(-1)^{4\boldsymbol{\varepsilon}\cdot \boldsymbol{\delta}}
\, e^{2\pi \iu\,\boldsymbol{\varepsilon} \cdot \mathcal{Q} \cdot \boldsymbol{\varepsilon}}\,
\Theta\bigg[\hspace{-.1cm} \begin{array}{c}
 \boldsymbol{0} \\  \boldsymbol{\delta} 
\end{array}\hspace{-.1cm} \bigg]
(\tilde{\tau})^2
=
\frac{1}{2^{n-1}}
\sum_{\boldsymbol{\delta}} 
m_n(\boldsymbol{\delta})\,
\Theta\bigg[\hspace{-.1cm} \begin{array}{c}
 \boldsymbol{0} \\  \boldsymbol{\delta} 
\end{array}\hspace{-.1cm} \bigg]
(\tilde{\tau})^2,
\end{equation}
where
\begin{equation} \label{eq:qdelta def}
m_n(\boldsymbol{\delta}) 
=
\sum_{\boldsymbol{\varepsilon}} 
(-1)^{4\boldsymbol{\varepsilon}\cdot \boldsymbol{\delta}} \,
e^{2\pi \iu\,\boldsymbol{\varepsilon} \cdot \mathcal{Q} \cdot \boldsymbol{\varepsilon}} 
= \sum_{\boldsymbol\varepsilon}e^{4 i\pi \left(\boldsymbol\varepsilon \cdot \frac{\mathcal Q}{2} \cdot \boldsymbol\varepsilon + \boldsymbol\varepsilon\cdot\boldsymbol\delta\right)}.
\end{equation}
In App.~\ref{app:sd boson} we show that $m_n(\boldsymbol\delta)$ can be written as the partition function of a classical Ising spin 
system, where $\boldsymbol\varepsilon$ play the role of the spin variables and the $\boldsymbol\delta$ are the local magnetic fields.
In the same appendix we also employ standard transfer matrix techniques to prove 
that $m_n(\boldsymbol\delta) = r_n(\boldsymbol\delta)$ (see (\ref{eq:r from EZ + correspondence}) and (\ref{eq:qdelta def})).

\subsection{Numerical checks}
\label{subsec:FF numerics}

\begin{figure}
\hspace{-1cm}
\includegraphics[width=1.08\textwidth]{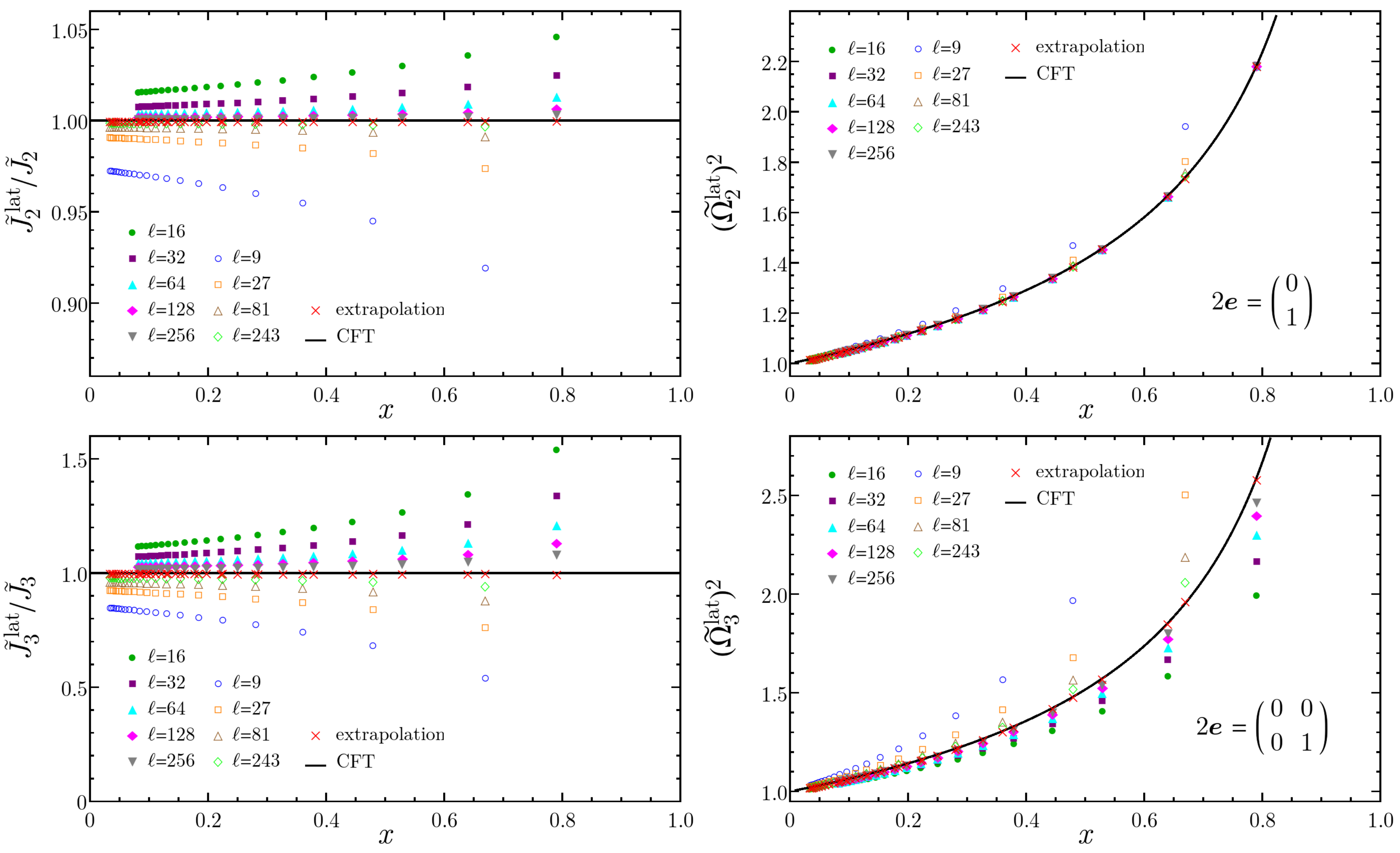}
\caption{
The terms occurring in $\Tr(\rho_A^{T_2})^n$ for the free fermion (see \eqref{eq:partial transpose free fermion}), according to the correspondence (\ref{eq:NegFF}).
Here we show $n=2$ (top panels) and $n=3$ (bottom panels).
For each group of identical terms, only one representative has been plotted.
In the left panels, the term with $\boldsymbol{p}=\boldsymbol{0}$ has been divided by its CFT counterpart ($\boldsymbol{\delta}=\boldsymbol{0}$), in order to simplify the residual dependence on $\ell_1$ and $\ell_2$.
The extrapolated points (red crosses) are obtained through a fit of the data according to the scaling function \eqref{eq:scaling Omega} and they agree with the CFT predictions (solid lines).
} 
\label{fig:FF n=2-3}
\end{figure}

\begin{figure}
\hspace{-1cm}
\includegraphics[width=1.08\textwidth]{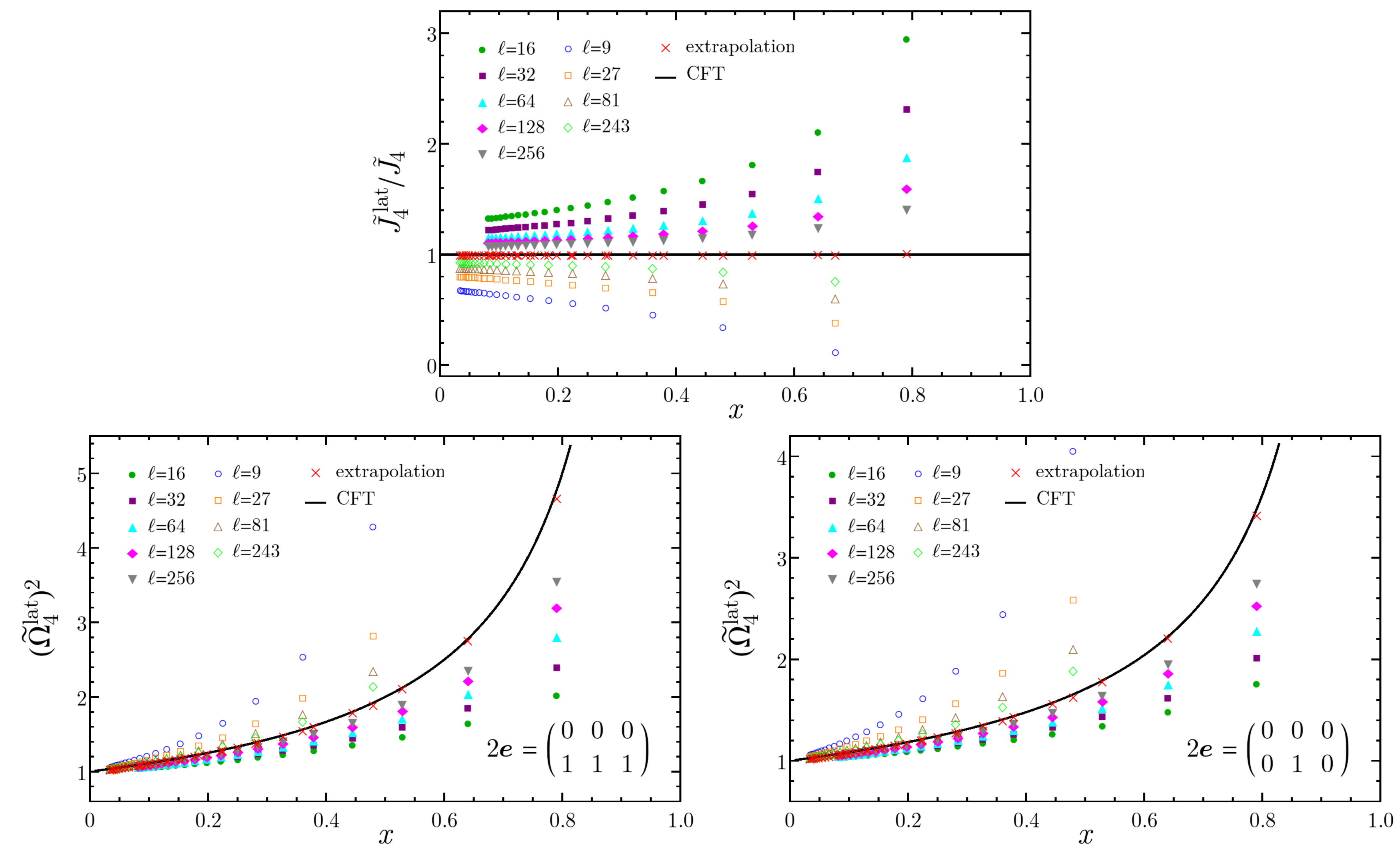}
\caption{
The terms occurring in $\Tr(\rho_A^{T_2})^4$ for the free fermion (see \eqref{eq:partial transpose free fermion}), according to the correspondence (\ref{eq:NegFF}).
For each of the three groups of identical terms, only one representative has been plotted.
In the top panel, the term with $\boldsymbol{p}=\boldsymbol{0}$ has been divided by its CFT counterpart ($\boldsymbol{\delta}=\boldsymbol{0}$), in order to simplify the residual dependence on $\ell_1$ and $\ell_2$.
The extrapolated points (red crosses) are obtained through a fit of the data according to the scaling function \eqref{eq:scaling Omega} and they agree with the CFT predictions (solid lines).
} 
\label{fig:FF n=4}
\end{figure}

In \cite{cct-15} we have already shown that $ \Tr ( \rho_A^{T_2})^n$ for $n=3,4,5$ converges in the continuum limit to the 
CFT predictions \eqref{eq:partial transpose free fermion}.
In this paper we have given a set of more stringent relations \eqref{eq:NegFF} between each  term in the sum
for $ \Tr ( \rho_A^{T_2})^n $ appearing both in CFT and on the lattice.
The goal of this subsection is to provide explicit numerical evidence of this term-by-term correspondence for $n=2,3,4$.

In order to evaluate numerically the traces of product of these matrices, we employ the techniques first developed in 
\cite{fc-10} for $\Tr \rho_A^n$ and recently used to compute $ \Tr ( \rho_A^{T_2})^n $ in \cite{cct-15}. 
Indeed, being the tight-binding  Hamiltonian \eqref{eq:Ham} quadratic in the fermionic operators, the ground state reduced density 
matrix $\rho_A$ is Gaussian.
Moreover, $\tilde\rho_A$ in (\ref{eq:ez}) is Gaussian as well and, since the string operator $P_C$ can be written as the exponential of a quadratic operator, the density matrix $P_{A_2}\tilde\rho_A P_{A_2}$ is also Gaussian.
Nevertheless, the sum of these two matrices in (\ref{eq:ez}) is not Gaussian (this is indeed the main difficulty compared to 
bosonic models in which the partial transpose is itself Gaussian \cite{aepw-02,br-04,pe-09,mrpr-09}).
By exploiting the fact that for Gaussian states all the information of the system is encoded in the correlation matrices, 
the computations can be performed in a polynomial time in terms of the total size of the subsystem.
In particular, in our case the correlation matrices of $\tilde\rho_A$ and $P_{A_2}\tilde\rho_A P_{A_2}$ can be obtained from the one of $\rho_A$, as described in \cite{ez-15,cct-15}.

The lattice computations have been performed in an infinite chain.
The disjoint blocks $A_1$ and $A_2$ have been taken with the same size $\ell_1 = \ell_2 \equiv \ell$, while the size of the block $B_1$ separating them is $d$.
Thus, the four point ratio (\ref{cross ratio def}) becomes
\begin{equation}
 x=\left(\frac{\ell}{\ell+d}\right)^2,
\end{equation}
and configurations with the same value of $\ell/d$ correspond to the same $x$.

Referring to Eq.~\eqref{pt A2 sum v1}, let us introduce the following lattice quantities
\be
 \tilde J_n^\text{lat} = 
  \Tr  \tilde{\rho}_{0}^n,
  \qquad
 \widetilde\Omega_n^\text{lat}[\boldsymbol p]^2  =  \frac{1}{\Tr  \tilde{\rho}_{0}^n}\;
 \Tr\bigg[ \tilde{\rho}_{0} \prod_{k=1}^{n-1}\tilde{\rho}_{p_k} \bigg] ,
\ee
which can be evaluated as explained in~\cite{cct-15}.
We also introduce their CFT continuum limit:
\be
\label{tilde Jn Omega def}
\tilde{J}_n
\equiv 
 c_n^2 \left( \frac{1-x}{\ell_1\ell_2} \right)^{2\Delta_n},
\qquad
\widetilde{\Omega}_n[ \boldsymbol e ]  
\equiv 
 \bigg|\frac{\Theta[ \boldsymbol e ] 
(\tilde\tau(x))}{\Theta(\tilde\tau(x))}\bigg|.
\ee
These CFT values are approached by taking configurations with increasing $\ell$, keeping the ratio $\ell/d$ fixed.
As discussed in Sec.~\ref{subsec dih symm renyi free}, many terms in the sum \eqref{pt A2 sum v1} are equal 
because of the properties of the trace (in the continuum, this degeneracy is due to the dihedral symmetry of the Riemann surface).

In order to deal with the finite size effects, we perform an accurate scaling analysis, as done in \cite{fc-10} for $ \Tr \rho_A^n $  and in \cite{cct-15} for $\Tr ( \rho_A^{T_2})^n$.
From general CFT arguments it has been shown that these quantities display some unusual corrections to the scaling in $\ell$ described by a power law term with exponent $\delta_n=2h/n$, being $h$ the smallest scaling dimension of a relevant operator inserted at the 
branch points \cite{cc-10,ccen-10,ce-10,un-vari}.
For the Dirac fermion $h=1$ and terms of the form $\ell^{-2m/n}$ are present, for any positive integer $m$.
Because of the slow convergence of these terms (which becomes slower and slower for increasing $n$), typically it is necessary to include in the scaling function many of them.
The most general finite-$\ell$ ansatz for $\widetilde \Omega_n$ takes the following form
  \begin{equation} 
  \label{eq:scaling Omega}
  \widetilde\Omega_n^\text{lat}[\boldsymbol p]^2 
  = \widetilde\Omega^2_n[\boldsymbol e] + \frac{\omega^{(1)}_n(x)}{\ell^{2/n}} + \frac{\omega^{(2)}_n(x)}{\ell^{4/n}} + \frac{\omega^{(3)}_n(x)}{\ell^{6/n}} + \dots,
  \end{equation}
where $\widetilde\Omega[\boldsymbol e]$ is defined in~\eqref{tilde Jn Omega def} and $\boldsymbol p$ and $\boldsymbol e$ are related through~\eqref{corr delta-g} and~\eqref{p_i inverse}.
For the $\tilde J_n^\text{lat}/\tilde J_n$, a scaling function similar to \eqref{eq:scaling Omega} can be studied. 
Fitting the data with \eqref{eq:scaling Omega}, the more terms we include, the more precise the fit could be.
Nevertheless, since we have access to limited values of $\ell$, by using too many terms overfitting problems may be encountered, which lead to very unstable results. 
The number of terms to be included in \eqref{eq:scaling Omega} has been chosen in order to get stable fits.
We find that every term $\widetilde\Omega_n[\boldsymbol e]$ follows the scaling \eqref{eq:scaling Omega} and the extrapolated value agrees with the corresponding CFT result.

Our numerical results are shown in Fig.~\ref{fig:FF n=2-3} for $n=2$ (top panels) and $n=3$ (bottom panels), while Fig.~\ref{fig:FF n=4} is about the $n=4$ case.
As for the prefactor, the ratio $\tilde J_n^\text{lat}/\tilde J_n$ has been considered in order to eliminate the trivial dependence on $\ell$ which survives in the continuum limit.
The solid lines are the CFT predictions, which are given by (\ref{eq:NegFF}).

\section{Conclusions}
\label{concl}

In this manuscript we studied the moments of the partial transpose of the reduced density matrix $\Tr(\rho_A^{T_2})^n$
for 
two disjoint intervals in the conformal field theory of the massless Dirac fermion. 
Our main result is a closed analytic form for these moments of arbitrary order, i.e.\ Eq.~\eqref{eq:partial transpose free fermion}.
For $n=3,4,5$ this formula was anticipated in Ref.~\cite{cct-15}, but we extend here to arbitrary $n$
and provide its full derivation. 
The analytic computation of the logarithmic negativity $\mathcal{E}$ through the replica limit of (\ref{eq:partial transpose free fermion}) for even $n_e \to 1$ is beyond our knowledge.

It turned out that these moments are identical, for arbitrary order, to those of the compactified boson 
at the self-dual point. 
This equality comes from the explicit computation and we miss a proper 
understanding of this fact. 
It was already noticed that for the moments of the reduced density matrix of two disjoint intervals $\Tr \rho_A^n$, 
the result for the free fermion \cite{ch-05,pascazio-08} and the one for the compactified boson at the self-dual radius \cite{cct-09}
are equal and very easy (they are both given by (\ref{Fn}) with ${\cal F}_n(x)=1$). 
This is not the case for three or more disjoint intervals \cite{hlr-13, ctt-14}.
This unexpected equivalence has been investigated in \cite{hlr-13}, where also other results have been found, 
based on the fact that $\tau(x)$ is purely imaginary when $x\in (0,1)$.
For the partial transpose, the period matrix $\tilde{\tau}(x)$ of $\widetilde{\mathcal{R}}_n$ in (\ref{eq:Q}) has a non vanishing real part. 
Nevertheless, here we have shown that $\Tr \left( \rho_A^{T_2}\right)^n$ for the free fermion is equal to the one for the self-dual boson, 
a property that does not follow from the analysis of \cite{hlr-13}. 
%
The equality of all the moments obviously implies also the equality of the negativities.
Since the negativity is directly measurable by means of tensor network algorithms 
(as e.g.\ done in \cite{wmb-09,ctt-13} for the Ising model), it would be very interesting to check numerically the 
identity between the negativity of the tight-binding model and the isotropic Heisenberg antiferromagnet (whose continuum limit is the 
self-dual boson). This is a highly non trivial prediction.

We point out that an interesting technical byproduct of this paper is a one-to-one correspondence between 
each of the $2^{n-1}$ terms appearing in the lattice formulation of the moments $\Tr(\rho_A^{T_2})^n$
and the partition function of the free fermion on $\widetilde{\mathcal{R}}_n$ with a particular assignment of 
fermionic boundary conditions. 
This correspondence has been explicitly checked against lattice numerical computations extending the numerical analysis 
of Ref.~\cite{cct-15} where only the overall sum was considered. 
The consequence of this correspondence for the moments of both the reduce density matrix and its partial transpose in 
spin models with a free fermionic representation will be explored elsewhere \cite{cct-prep}.

\section*{Acknowledgments}

We are very grateful to Maurizio Fagotti for the collaboration at an early stage of this project. 
We are grateful in particular to Horacio Casini and Tamara Grava for important discussions. 
We thank Matthew Headrick, Christopher Herzog,  Albion Lawrence, Israel Klich and Yihong Wang for useful comments or correspondence.
AC and ET thank Kavli Institute for Theoretical Physics for warm hospitality, financial support and a stimulating environment during the program {\it Entanglement in Strongly-Correlated Quantum Matter} where part of this work has been done.
For the same reasons, ET and PC are also grateful to Galileo Galilei Institute for the 
program {\it Holographic Methods for Strongly Coupled Systems}. 
PC and ET have been supported by the ERC under Starting Grant  279391 EDEQS.

\begin{appendices}

\section*{Appendices}

\section{Reduced density matrix and its partial transpose on the lattice}
\label{app: rdm}

In this appendix we briefly review the main tools employed in this manuscript to study the partial transpose 
for free fermions on the lattice.

Given the reduced density matrix (\ref{eq:rhoA}), it is convenient to distinguish the terms having an even or odd number of fermionic operators in $A_2$ (notice that the parity of operator in $A_2$ is the same of the operators in $A_1$) by introducing
\begin{equation}
\label{rhoA even-odd def}
 \rho_\text{even} = \frac{1}{2^{\ell_1+\ell_2}}\sum_\text{even}w_{12}\,O_1 O_2,
 \qquad
 \rho_\text{odd}  = \frac{1}{2^{\ell_1+\ell_2}}\sum_\text{odd}w_{12}\,O_1 O_2.
\end{equation}
Thus $ \rho_A =  \rho_\text{even}  +  \rho_\text{odd}$.
The partial transposition with respect to $A_2$ in (\ref{hat transposition def}) acts differently on the two operators in (\ref{rhoA even-odd def}). 
In particular \cite{ez-15}
\begin{equation}
 \rho^{T_2}_\text{even} = \frac{1}{2^{\ell_1+\ell_2}}\sum_\text{even}(-1)^{\mu_2/2}w_{12}\,O_1 O_2,
 \qquad
 \rho^{T_2}_\text{odd}  = \frac{1}{2^{\ell_1+\ell_2}}\sum_\text{odd} (-1)^{(\mu_2-1)/2}w_{12}\,O_1 O_2.  
\end{equation}
Defining the following Gaussian matrix $\tilde\rho_A$ (which is not a density matrix, being not hermitian)
\begin{equation}
\label{rhoAtilde O1O2}
 \tilde\rho_A = \frac{1}{2^{\ell_1+\ell_2}} \sum \iu^{\mu_2} w_{12}\,O_1 O_2,
\end{equation}
and also $ \tilde\rho_\text{even} $ and $ \tilde\rho_\text{odd} $ as done in (\ref{rhoA even-odd def}) for $\rho_A$,
the partial transpose of $\rho_A$ becomes
\begin{equation} 
\label{eq:rhoT even odd}
 \rho_A^{T_2} = \tilde\rho_\text{even} - \iu \, \tilde\rho_\text{odd}.
\end{equation}
The matrices $ \tilde\rho_\text{even} $ and $ \tilde\rho_\text{odd} $ in (\ref{eq:rhoT even odd}) can be written 
through the string $P_{A_2}$ of the Majorana operator along $A_2$.
Since $P_{A_2} a_j^{x,y} P_{A_2} = (-1)^{\delta_{j\in A_2}} \, a_j^{x,y}$, one finds
\begin{equation}
\label{rho tilde even-odd}
 \tilde{\rho}_\text{even} = \frac{1}{2}\big( \tilde{\rho}_A + P_{A_2} \tilde{\rho}_A P_{A_2} \big),
 \qquad\quad
 \tilde{\rho}_\text{odd}  = \frac{1}{2}\big( \tilde{\rho}_A - P_{A_2} \tilde{\rho}_A P_{A_2} \big).
\end{equation}
Thus, plugging (\ref{rho tilde even-odd}) into (\ref{eq:rhoT even odd}), the final expression for $ \rho_A^{T_2} $ in (\ref{eq:ez}) is obtained \cite{ez-15}.

\section{A check for $n=2$}
\label{app check n=2}

In this appendix, by employing the formalism described in Sec.~\ref{subsec coherent state site} and Sec.~\ref{subsec pt coherent}, we check the standard relation between the reduced density matrix of two disjoint intervals and its partial transpose
\be
\Tr\left( \rho_A^{T_2} \right)^2 = \Tr\left( \rho_A^2 \right).
\ee
From (\ref{eq:ez}), it is immediate to observe that $\Tr\left( \rho_A^{T_2} \right)^2 = \Tr\left( \tilde\rho_A P_{A_2} \tilde\rho_A P_{A_2} \right)$. 
Then, by using \eqref{eq:identity and trace}, we can write
\bea
 & &  \hspace{-2.3cm}
 \Tr\left( \tilde\rho_A P_{A_2} \tilde\rho_A P_{A_2} \right) \,=\, \\
 & & \hspace{-1.8cm} =\,
 \int D\conj{\zeta} D\zeta\,D\conj{\eta} D\eta\, e^{-\conj\zeta\zeta}e^{-\conj\eta\eta}
 \braket{-\zeta_1,\,-\zeta_2|\tilde\rho_A|\eta_1,\,\eta_2}\braket{\eta_1,\,\eta_2|P_{A_2}\tilde\rho_A P_{A_2}|\zeta_1,\,\zeta_2} \nonumber \\
 & &  \hspace{-1.8cm} =\,
 \int D\conj{\zeta} D\zeta\,D\conj{\eta} D\eta\, e^{-\conj\zeta\zeta}e^{-\conj\eta\eta}
 \braket{-\zeta_1,\,-\zeta_2|\tilde\rho_A|\eta_1,\,\eta_2}\braket{\eta_1,\,-\eta_2|\tilde\rho_A|\zeta_1,\,-\zeta_2} \nonumber \\
 & &  \hspace{-1.8cm} =\,
  \int D\conj{\zeta} D\zeta\,D\conj{\eta} D\eta\, e^{-\conj\zeta\zeta}e^{-\conj\eta\eta}
 \braket{-\zeta_1,\,\conj\eta_2|\rho_A|\eta_1,\,\conj\zeta_2}\braket{\eta_1,\,-\conj\zeta_2|\rho_A|\zeta_1,\,\conj\eta_2} \nonumber \\
 \label{eq:Trrho2}
 & &  \hspace{-1.8cm} =\,
  \int D\conj{\zeta} D\zeta\,D\conj{\eta} D\eta\, e^{-\conj\zeta_1\zeta_1+\conj\zeta_2\zeta_2}e^{-\conj\eta_1\eta_1+\conj\eta_2\eta_2}
  \braket{-\zeta_1,\,\eta_2|\rho_A|\eta_1,\,\zeta_2}\braket{\eta_1,\,-\zeta_2|\rho_A|\zeta_1,\,\eta_2} \nonumber,
\eea
where in the last step the change of variables $\zeta_2 \rightarrow \conj\zeta_2$, $\eta_2 \rightarrow \conj\eta_2$ has been employed. 
Then, by noticing that the relations in (\ref{eq:identity and trace}) can be slightly modified as follows
\begin{equation} \label{eq:identity and trace 2}
 -\mathbb{I} = \intf{\zeta}e^{\conj\zeta\zeta}\ket{\zeta}\bra{-\zeta}, 
 \qquad 
 - \, \Tr{\,\hat O} = \intf{\zeta}e^{\conj\zeta\zeta}\bra{\zeta}\hat O\ket{\zeta} ,
\end{equation}
we can conclude that \eqref{eq:Trrho2} is exactly $\Tr\left(\rho_A^2\right)$.

\section{On the Riemann surface $\widetilde{\mathcal R}_n$}
\label{app:Rn}

In this appendix we derive with all the details some results about the Riemann surface $\widetilde{\mathcal R}_n$ and its period 
matrix $\tilde{\tau}(x)$ in (\ref{eq:Q}) that are employed in the main text.

Given the two disjoint intervals $A_1 =(u_1,v_1)$ and $A_2 =(u_2,v_2)$ whose endpoints are ordered as $u_1 < v_1 < u_2 < v_2$, 
$\Tr \rho_A^n$ is the partition function of the model on the Riemann surface $\mathcal R_n$ which is defined by the following 
algebraic curve in $\mathbb{C}^2$ (parameterised by the complex variables $z$ and $y$) \cite{cct-09}
\begin{equation} 
\label{eq:Rn}
 y^n = (z-u_1)(z-u_2)\big[(z-v_1)(z-v_2)\big]^{n-1} .
\end{equation}
The Riemann surface $\mathcal R_n$, which is an $n$ sheeted cover of the complex plane, has genus $n-1$ and it has been studied in detail in \cite{eg-03, ctt-14}, where its generalisation to any number of disjoint intervals (whose genus is $(n-1)(N-1)$ for $N$ intervals) has been also discussed. 
By going around $u_1$ and $u_2$ clockwise, one goes from the $j$-th to the $(j+1)$-th sheet, while going around 
$v_1$ and $v_2$ clockwise, one moves to the $(j-1)$-th one.

In order to compute $\Tr ( \rho_A^{T_2} )^n $, one has to find the partition function of the model on a different Riemann surface $\widetilde{\mathcal R}_n$, which is obtained by exchanging $u_2\leftrightarrow v_2$ in \eqref{eq:Rn}, i.e.
\begin{equation} 
\label{eq:Rtilden}
 y^n = (z-u_1)(z-v_2)\big[(z-v_1)(z-u_2)\big]^{n-1} .
\end{equation}
The Riemann surface $\widetilde{\mathcal R}_n$ has genus $n-1$ and the sheets are joined in a different way with respect to 
$\mathcal{R}_n$. Indeed, by encircling $u_2$ clockwise we move from the $j$-th to the $(j-1)$-th sheet while by encircling 
$v_2$ clockwise the $(j+1)$-th sheet is reached.

While the period matrix $\tau(x)$ of $\mathcal{R}_n$ is purely imaginary (see (\ref{eq:tau})), the period matrix $\tilde{\tau}(x)$ for $\widetilde{\mathcal R}_n$ has a non vanishing real part. 
In Sec.~\ref{app:Q} we show that $\textrm{Re}[\tilde{\tau}(x)]$ has a very simple form. 
In Sec.~\ref{app: small x} we consider the moments $ \Tr\left(\rho_A^{T_2}\right)^n $ in the regime of small intervals $x \to 0$ and in Sec.~\ref{app:dihn} we provide a detailed discussion of the $\mathbb{Z}_2$ part of the dihedral symmetry for $\widetilde{\mathcal{R}}_n$.

\subsection{The real and imaginary part of the period matrix $\tilde{\tau}(x)$}
\label{app:Q}

In this subsection we want to write explicitly  the real and the imaginary part of the period matrix $\tilde\tau(x)$ given by (\ref{eq:Q}).
The real part $\textrm{Re}[\tilde{\tau}(x)]$ turns out to be a simple tridiagonal matrix with half-integer entries.

Let us introduce the following ratios of hypergeometric functions, which enter in the expressions for the period matrices $\tau(x)$ and $\tilde{\tau}(x)$ (see  (\ref{eq:tau})  and (\ref{eq:Q}))
\begin{equation}
\tau_{r}(x) \equiv \iu\, \frac{{}_2F_1(r,1-r;1;1-x) }{{}_2F_1(r,1-r;1;x) } ,
\qquad\quad
\tilde{\tau}_{r}(x) \equiv \tau_{r}\Big( \frac{x}{x-1} \Big) \equiv \tilde\alpha_r(x) + \iu \tilde\beta_r(x),
\end{equation}
where $0<r<1$ and $x\in (0,1)$.
Moreover, we also define
\bea
\label{Ar def}
\mathcal{A}_{r}(x) &\equiv& \frac{\Gamma(1-2r)}{\Gamma(1-r)^2} \; \frac{_2F_1(r,r;2r;1-x)}{_2F_1(r,r;1;x)}, \\
\rule{0pt}{.7cm}
\label{Br def}
\mathcal{B}_{r}(x) &\equiv &\frac{\Gamma(2r-1)}{\Gamma(r)^2} \,(1-x)^{1-2r} \,\frac{_2F_1(1-r,1-r;2(1-r);1-x)}{_2F_1(r,r;1;x)}.
\eea
By employing the expressions given in (87) of \cite{cct-neg-long}, one finds that
\be
\label{eq:taukn}
\tilde\alpha_{r}(x) \,=\, \sin(\pi r) \big[ \mathcal{A}_{r}(x)  + \mathcal{B}_{r}(x) \big] ,
\qquad
\tilde\beta_{r}(x) \,=\, \cos(\pi r) \big[ \mathcal{A}_{r}(x)  - \mathcal{B}_{r}(x) \big].
\ee
At this point we need the following identity (see e.g.\ Eq.~(1) at pag.~108 of Ref.~\cite{bateman})
\bea
\label{2F1 id1}
& & \hspace{-2.5cm}
_2F_1(a,b;c;z) \,=\,
\frac{\Gamma(c)\,\Gamma(c-a-b)}{\Gamma(c-a)\,\Gamma(c-b)}\, _2F_1(a,b;a+b-c+1;1-z) 
\\
\rule{0pt}{.65cm}
& & \hspace{.5cm}
+  \frac{\Gamma(c)\,\Gamma(a+b-c)}{\Gamma(a)\,\Gamma(b)}\,(1-z)^{c-a-b}\, _2F_1(c-a,c-b;c-a-b+1;1-z),
\nonumber
\eea
which holds for $|\textrm{arg}(1-z)| < \pi$.
By specialising  (\ref{2F1 id1}) to the case of $(a,b,c)=(r,r,1)$ and $z=x\in (0,1)$, from (\ref{Ar def}) and (\ref{Br def})  one finds that
\begin{equation}
\label{A+B=1}
\mathcal{A}_{r}(x)  + \mathcal{B}_{r}(x)  = 1.
\end{equation}
From this expression it is clear that the $x$ dependence disappears from the real part of $\tilde\tau_r(x)$ and hence from the period matrix.
Indeed, using~\eqref{eq:taukn} and~\eqref{A+B=1} in~\eqref{eq:tau} one gets (see also (\ref{eq:Q}))
\begin{equation}  
\label{eq:tau2inv reimparts}
\tilde{\tau}(x)_{i,j} 
\,=\, 
 \frac{2}{n} \sum_{k=1}^{n-1} \sin(\pi k/n) \, \tilde{\tau}_{k/n}(x)\,  \cos[2\pi (k/n)(i-j)]
\,=\,
\frac{1}{2}\,\big[\mathcal{Q}\big]_{i,j}  + \iu\,\big[\mathcal{I}(x)\big]_{i,j} ,
\end{equation}
where the sum giving the real part can be explicitly performed, finding that the matrix $\mathcal{Q}$ has integer elements which read
\begin{equation}
\label{Qmat explicit}
[\mathcal{Q}]_{i,j} \equiv  2 \delta_{i,j} - \delta_{|i-j|,1} ,
\end{equation}
namely $\mathcal{Q}$ is a symmetric tridiagonal matrix, with $2$ on the main diagonal and $-1$ on the first diagonals.
On the other hand, the imaginary part can be written as follows,
\be
\label{bn matrix def}
 \big[\mathcal{I}(x)\big]_{i,j}  \,=\,
\frac{2}{n} \sum_{k=1}^{n-1} \sin(\pi k/n) \, 
\tilde{\beta}_{k/n}(x)
 \cos[2\pi (k/n)(i-j)] ,
\ee
with 
\be
\label{tilde beta def}
\tilde{\beta}_{r}(x) \equiv
\frac{\tilde{f}_r(1-x) - \tilde{f}_{1-r}(1-x) }{(1-x)^r\, _2F_1(r,r;1;x)} \, \cos(\pi r) ,
\qquad
\tilde{f}_{r}(x) \equiv  \frac{\Gamma(1-2r)}{\Gamma(1-r)^2}\; x^r \, _2F_1(r,r;2r;x) .
\ee

This expression for $\mathcal I(x)$ can be further simplified.
Plugging (\ref{A+B=1}) into the second expression of (\ref{eq:taukn}), we have
\begin{equation}
\label{eq:Im tau r}
\tilde\beta_r(x)  = \cos(\pi r) \big[ 2\mathcal{A}_{r}(x)  - 1\big] .
\end{equation}
For $0<x<1$ we can rewrite $\mathcal A_r(x) $ as follows \cite{bateman}
\begin{equation} \label{eq:A redef 2}
\mathcal A_r(x) =
 \frac{1}{2\cos(\pi r)}\left[ e^{- \iu\pi r} + e^{\iu\pi r} \,x^{-r} \,  \frac{_2F_1(r,r;1;1/x)}{_2F_1(r,r;1;x)}\,\right] .
\end{equation}
Since $\mathcal A_r(x)$ is real for $0<x<1$, the vanishing of its imaginary part gives
\begin{equation} \label{eq:impart id}
\frac{1}{x^r}\left[\tan(\pi r) \,\textrm{Re}\bigg( \frac{_2F_1(r,r;1;1/x)}{_2F_1(r,r;1;x)} \bigg)
+ \textrm{Im}\bigg( \frac{_2F_1(r,r;1;1/x)}{_2F_1(r,r;1;x)} \bigg) \right]
= \tan(\pi r) .
\end{equation}
On the other hand, by writing $\mathcal A_r(x)$ as its real part
and plugging the resulting expression in~\eqref{eq:Im tau r}, one finds
\begin{equation}
 \tilde\beta_r(x) = \frac{1}{x^r}
 \left[\cos(\pi r) \, \textrm{Re}\bigg( \frac{_2F_1(r,r;1;1/x)}{_2F_1(r,r;1;x)} \bigg) - \sin(\pi r) \, \textrm{Im}\bigg( \frac{_2F_1(r,r;1;1/x)}{_2F_1(r,r;1;x)} \bigg)\right] .
\end{equation}
Finally, using~\eqref{eq:impart id} we can write
\begin{equation} \label{eq:Im tau final}
 \tilde\beta_r(x) = \cos(\pi r) - \frac{x^{-r}}{\sin(\pi r) } \; \textrm{Im}\bigg( \frac{_2F_1(r,r;1;1/x)}{_2F_1(r,r;1;x)} \bigg) .
\end{equation}
The matrix $ \mathcal{I} (x)$ can be easily written by plugging \eqref{eq:Im tau final} into \eqref{bn matrix def} and noticing that the sum over the cosine vanishes.
The result reads
\be
\label{eq:tau2 inversion}
 \big[\mathcal{I}(x)\big]_{i,j} \,=\, 
 -\,  \frac{2}{n} \sum_{k=1}^{n-1} x^{-k/n} \; 
 \textrm{Im}\bigg( \frac{_2F_1(k/n,k/n;1;1/x)}{_2F_1(k/n,k/n;1;x)} \bigg)
\,\cos[2\pi (k/n)(i-j)].
\ee
The result (\ref{Qmat explicit}) is employed in Sec.~\ref{subsec:boson eta1} and in App.~\ref{app:dihn}.

\subsection{Short intervals regime}
\label{app: small x}

In this appendix we study the $ \Tr\left(\rho_A^{T_2}\right)^n $ for the free fermion \eqref{eq:partial transpose free fermion} in the limit of 
short intervals, i.e.\ when $x \to 0$.

In the expression \eqref{eq:partial transpose free fermion}, only Riemann theta functions with $\boldsymbol{\varepsilon} = \boldsymbol{0}$ occur, which are given by
\be
\label{theta eps=0}
\Theta 
\bigg[ \hspace{-.1cm} \begin{array}{cc} \boldsymbol{0} \\ \boldsymbol{\delta} \end{array} \hspace{-.1cm}\bigg]
\big(\tilde{\tau}(x)\big)
=
1+ \sum_{\boldsymbol{m}  \neq \boldsymbol{0} }
e^{\iu\pi (\boldsymbol{m}  \cdot \,\mathcal{Q} \cdot \,\boldsymbol{m}
+ 2 \boldsymbol{\delta} \cdot \boldsymbol{m}
)}
\,
e^{- \pi \, \boldsymbol{m}  \cdot \,\mathcal{I} \cdot \,\boldsymbol{m}},
\ee
where $\mathcal{Q} $ is independent of $x$.
Expanding $\tilde{\beta}_{k/n}(x)$ in (\ref{tilde beta def}) for $x\to 0$, one finds
\be
\tilde{\beta}_{q}(x) = -\frac{\sin(\pi q)}{\pi}\,
\big[ 
\log(x) + 2\gamma_{E} +\psi(q) + \psi(1-q) 
\big]
+O(x).
\ee
Plugging this expansion into (\ref{bn matrix def}) and (\ref{theta eps=0}), one gets that the leading term is $x^{\boldsymbol{m}  \cdot \,\mathcal{Q} \cdot \,\boldsymbol{m}}$.
The exponent $\boldsymbol{m}  \cdot \mathcal{Q} \cdot \boldsymbol{m}$ for $\boldsymbol{m} \in \mathbb{Z}^{n-1}$ has been already analyzed in \cite{cct-11}, finding that its minimum is $1$, which is obtained for the following vectors
\be
\boldsymbol{m}_\pm \equiv 
\big( \overbrace{\rule{0pt}{.35cm}\underbrace{0\,, \dots , \, 0}_{j_1} \,, \pm 1\,, \dots , \pm 1}^{j_2}\,, \, 0\,, \dots, \, 0\, \big),
\qquad
\boldsymbol{m}_\pm \neq \boldsymbol{0},
\qquad
0 \leqslant j_1 < j_2 \leqslant n-1,
\ee
namely $\boldsymbol{m}_\pm  \cdot \mathcal{Q} \cdot \boldsymbol{m}_\pm =1$.
Then, by applying again the results of \cite{cct-11} (notice that the vectors $\boldsymbol{m}_+$ and $\boldsymbol{m}_-$ 
give the same contribution) to (\ref{theta eps=0}), we find
\be
\Theta 
\bigg[ \hspace{-.1cm} \begin{array}{cc} \boldsymbol{0} \\ \boldsymbol{\delta} \end{array} \hspace{-.1cm}\bigg]
\big(\tilde{\tau}(x)\big)
\,=\,
1- \frac{x}{2n^2} 
\sum_{\boldsymbol{m}_+}
\frac{(-1)^{2 \boldsymbol{\delta} \cdot \boldsymbol{m}_+}}{\sin^2(\pi(j_2-j_1)/n)}
+ \dots,
\ee
where the dots denote $o(x)$ terms. 
Thus, for the generic term occurring in the sum \eqref{eq:partial transpose free fermion} we have
\be
\label{tilde Omega x=0}
\widetilde{\Omega}_n[ \boldsymbol e ]  
\,=\,
1
-\frac{x}{2n^2}
\sum_{\boldsymbol{m}_+} 
\frac{1+(-1)^{2\boldsymbol{\delta} \cdot \boldsymbol{m}_+}}{\sin^2(\pi (j_2-j_1)/n)}
+\dots,
\qquad
\boldsymbol e 
=
\bigg( \hspace{-.1cm} \begin{array}{cc} \boldsymbol{0} \\ \boldsymbol{\delta} \end{array} \hspace{-.1cm}\bigg).
\ee
Plugging this result into \eqref{eq:partial transpose free fermion}, we get the first term of the $x\to 0$ expansion of $\Tr\left(\rho_A^{T_2}\right)^n$, which is 
\be
 \Tr\left(\rho_A^{T_2}\right)^n 
 =  
 \, \tilde{J}_n \bigg[\, 1 
 -\frac{x}{2^{n-1}\, n^2} \sum_{\boldsymbol{\delta}}  r_n(\boldsymbol\delta)\,
\sum_{\boldsymbol{m}_+} 
\frac{1+(-1)^{2\boldsymbol{\delta} \cdot \boldsymbol{m}_+}}{\sin^2(\pi (j_2-j_1)/n)}
+ \dots
 \bigg].
\ee

\subsection{The $\mathbb{Z}_2$ part of the dihedral symmetry of $\widetilde{\mathcal{R}}_n$}
\label{app:dihn}

In this subsection we briefly discuss the most peculiar aspect of the dihedral symmetry for the Riemann surface $\widetilde{\mathcal{R}}_n$ occurring in the computation of $ \Tr\left(\rho_A^{T_2}\right)^n $ (see Sec.~\ref{subsec dih symm renyi free}).

The Riemann surface $\mathcal{R}_n$ has a dihedral symmetry $\mathbb{Z}_n\times \mathbb{Z}_2$ due to the invariance under cyclic permutation of the sheets ($\mathbb{Z}_n$) and the complex conjugation ($\mathbb{Z}_2$).
For a genus $g$ Riemann surface, the modular transformations are given by the symplectic matrices $Sp(2g,\mathbb{Z})$ \cite{bosonization higher genus, dvv}. 
The dihedral symmetry can be identified with a subgroup of the modular transformations acting on $\mathcal{R}_n$ which has been discussed in \cite{hlr-13,ctt-14}.
In particular, these peculiar modular transformations map the $a$ cycles among themselves and the $b$ cycles among themselves, leaving the period matrix $\tau(x)$ unchanged.

Also the surface $\widetilde{\mathcal{R}}_n$ has a dihedral symmetry $\mathbb{Z}_n\times \mathbb{Z}_2$ but, while the cyclic permutation ($\mathbb{Z}_n$) is exactly the same one discussed above for $\mathcal{R}_n$, the complex conjugation is slightly different because it mixes $a$ and $b$ cycles. 
Let us remind that the complex conjugation corresponds to reverse the order of the sheets and to reflect all of them with respect to the real axis. 
Considering the canonical homology basis $\{\tilde{a}_r , \tilde{b}_r ; 1\leqslant r \leqslant n-1\}$ introduced in Sec.~\ref{sec cont limit free} (see Fig.~\ref{fig:cycles b}) \cite{ctt-14}, this transformation acts as follows
\begin{equation}
\label{M_inv tilde def}
\bigg( \hspace{-.1cm} \begin{array}{cc} 
\tilde{\boldsymbol{a}}' \\ \tilde{\boldsymbol{b}}'
\end{array} \hspace{-.1cm}\bigg)
 =
 M_\text{inv}
  \cdot
\bigg( \hspace{-.1cm} \begin{array}{cc} 
\tilde{\boldsymbol{a}} \\ \tilde{\boldsymbol{b}}
\end{array} \hspace{-.1cm}\bigg),
 \qquad\quad
 M_\text{inv} = 
 \begin{pmatrix}
\, \overset{\leftrightarrow}{\mathbb{I}}_{n-1} & 0 \\ 
 - \overset{\leftrightarrow}{\mathcal Q} & \overset{\leftrightarrow}{\mathbb{I}}_{n-1}\, 
 \end{pmatrix}
 \in 
\, Sp\big(2(n-1),\mathbb{Z}\big),
\end{equation}
where we introduced the notation of the double-headed arrow above a matrix to indicate that the columns have to be taken in the reversed order
($\mathbb{I}_{k}$ is the $k \times k$ identity matrix and $\mathcal Q$ is given by (\ref{Qmat explicit})).
Under the symplectic transformation defined in (\ref{M_inv tilde def}), the period matrix $\tilde\tau$ (\ref{eq:Q}) changes as follows
\be
\label{tau tilde prime}
 \tilde\tau' = \overset{\leftrightarrow}{\mathbb{I}}\cdot \tilde\tau \cdot \overset{\leftrightarrow}{\mathbb{I}} - \overset{\leftrightarrow}{\mathcal Q}\cdot \overset{\leftrightarrow}{\mathbb{I}}
 = \tilde\tau - \mathcal{Q} = -\mathcal R + \iu\, \mathcal I = -\,\conj{\tilde\tau} ,
 \ee
 while for the characteristics of the Riemann theta functions we have
\be
 \bigg( \hspace{-.1cm} \begin{array}{cc} 
  \boldsymbol \varepsilon' \\ \boldsymbol \delta'
\end{array} \hspace{-.1cm}\bigg)
 =
 \begin{pmatrix}
\, \overset{\leftrightarrow}{\mathbb{I}} & 0 \\ 
 \overset{\leftrightarrow}{\mathcal Q} & \overset{\leftrightarrow}{\mathbb{I}}
 \end{pmatrix}
  \cdot
 \bigg( \hspace{-.1cm} \begin{array}{cc} 
  \boldsymbol \varepsilon \\ \boldsymbol \delta
\end{array} \hspace{-.1cm}\bigg) .
\ee
Notice that the powers of $M_{\text{inv}}$ read
\begin{equation}
 M_{\text{inv}}^{2k-1} = 
 \begin{pmatrix}
 \overset{\leftrightarrow}{\mathbb{I}} & 0 \\ 
 \,- (2k-1)\overset{\leftrightarrow}{\mathcal Q} & \overset{\leftrightarrow}{\mathbb{I}}\,
 \end{pmatrix} ,
 \qquad
 M_{\text{inv}}^{2k} = 
 \begin{pmatrix}
 \mathbb{I} & 0 \\ 
 - 2k \mathcal Q & \mathbb{I}\,
 \end{pmatrix} ,
\end{equation}
(in particular, notice that $M_\text{inv}^2\neq\mathbb{I}$) so that, by applying (\ref{tau tilde prime}) $k$ times one finds
\begin{equation}
 \tilde\tau^{(k)} = \tilde\tau - k\,\mathcal{Q} .
\end{equation}
As for the change of the Riemann theta function under the modular transformation in (\ref{M_inv tilde def}), because of the particular form of $\mathcal Q$, it is easy to show that for $k$ even it is left invariant (a part for an overall sign), while
for $k$ odd it becomes its complex conjugate, up to an overall sign.
Since in our formulas the modulus of the Riemann theta function always occurs, the terms occurring in our sum over the characteristics are invariant under this transformation. Thus, $M_\text{inv}$ can be the modular transformation representing the $\mathbb{Z}_2$ of the dihedral symmetry, even if $M_\text{inv}^2\neq\mathbb{I}$.

\section{Details on the computation for the self-dual boson}
\label{app:sd boson}

In this appendix we show the equality $m_n(\boldsymbol\delta) = r_n(\boldsymbol\delta)$ between the coefficient $m_n(\boldsymbol\delta)$ in \eqref{eq:qdelta def}, coming from the self-dual boson approach, and the coefficient $r_n(\boldsymbol\delta)$ in \eqref{eq:r from EZ + correspondence} occurring in the expression obtained through the free fermion analysis.

Considering $m_n(\boldsymbol\delta)$ in \eqref{eq:qdelta def}, the expression in the exponent can be written as follows
\begin{equation}
  \boldsymbol\varepsilon \cdot \frac{\mathcal Q}{2} \cdot \boldsymbol\varepsilon + \boldsymbol\varepsilon\cdot\boldsymbol\delta 
  =
  - \sum_{i=1}^{n-2} \varepsilon_i \, \varepsilon_{i+1} + \sum_{i=1}^{n-1} \varepsilon_i^2 + \sum_{i=1}^{n-1} \varepsilon_i\delta_i .
\end{equation}
Then, defining the spin variables $\sigma_i=4\varepsilon_i - 1 = \pm 1$ and local magnetic fields $h_i = 4\delta_i$, we find  that $m_n(\boldsymbol\delta)$ is equal to the partition function $Z_n$ of $n-1$ Ising spins in a binary magnetic field $h_i=0,2$ (a part for the first and last site), which reads
\begin{equation}
\label{Zn sigma}
 Z_n = e^{\frac{\iu\pi}{4}\left( \sum_i h_i + n \right)}
 \sum_{\boldsymbol\sigma}
 \exp\bigg[ \,\iu\, \frac{\pi}{4}
  \bigg(- \sum_{i=1}^{n-2} \sigma_i  \sigma_{i+1} + \sum_{i=1}^{n-1} \sigma_i h_i + \sigma_1 + \sigma_{n-1} \bigg)\bigg].
\end{equation}
Given this Ising model representation, we can compute $m_n(\boldsymbol\delta)$ through standard transfer matrix techniques.
Following \cite{derrida}, let us introduce the conditioned partition function with the last spin set to $\mu$, namely
\begin{equation}
 Z_n(\mu) = e^{\frac{\iu\pi}{4}\left( \sum_i h_i + n \right)}
 \sum_{\boldsymbol\sigma}\exp\bigg[ \frac{\iu\pi}{4}
  \bigg(- \sum_{i=1}^{n-3} \sigma_i \, \sigma_{i+1} - \mu \,\sigma_{n-2} + \sum_{i=1}^{n-2} \sigma_i h_i 
  + \mu \,h_{n-1} + \sigma_1 + \mu \bigg)\bigg].
\end{equation}
Then, by adding one spin $\epsilon$ to the partition function, it becomes
\bea
 & & 
 \hspace{-2.5cm}
 Z_{n+1}(\epsilon) = e^{\frac{\iu\pi}{4}\left( \sum_i h_i + n \right)} \times
\\
& &
 \hspace{-2.3cm}
 \times \sum_{\mu=\pm 1}\sum_{\boldsymbol\sigma}\exp
 \bigg[ \frac{\iu\pi}{4}
  \bigg(- \sum_{i=1}^{n-3} \sigma_i\sigma_{i+1} - \mu\,\sigma_{n-2} -\mu\,\epsilon 
  + \sum_{i=1}^{n-2} \sigma_i\,h_i + \mu \,h_{n-1} + \epsilon \,h_{n} + \sigma_1 + \epsilon \bigg)\bigg].
  \nonumber
\eea
After some algebra, one realizes that
\begin{equation}
  Z_{n+1}(+) = (-1)^{2\delta_{n}} \left[ Z_{n}(+) - Z_{n}(-) \right] ,
  \qquad
  Z_{n+1}(-) = Z_{n}(+) + Z_{n}(-) .
\end{equation}
and these relations give the transfer matrix  
\begin{equation}
\label{hat T def}
 \widehat T(\delta) = \frac{1}{\sqrt 2}
 \begin{pmatrix}
  \, (-1)^{2\delta} & -(-1)^{2\delta} \, \\
  1 & 1
 \end{pmatrix} ,
 \qquad
 \delta \in \big\{0, 1/2\big\}.
\end{equation}
We also need the conditioned partition functions for a single spin, which read
\begin{equation}
 Z_2(+) = -\, e^{\frac{\iu\pi}{2}h_1},
 \qquad\quad 
 Z_2(-) = 1.
\end{equation}
Then, the partition function for $n-1$ spins in (\ref{Zn sigma}) is given by 
\bea
\label{Zn pm}
& & \hspace{-2cm}
 Z_n = Z_n(+) + Z_n(-) = 
 \\  \hspace{-2cm}
 & & 
 \hspace{-1cm}=\;
 2^{\frac{n}{2}-1} \,
 \big( \hspace{-.05cm} \begin{array}{cc}  1 & 1  \end{array} \hspace{-.05cm} \big)
 \prod_{k=n-1}^2
 \widehat T(\delta_k)\, 
 \bigg( \hspace{-.1cm} \begin{array}{c} 
  -(-1)^{2\delta_1} \\ 1
  \end{array} \hspace{-.1cm} \bigg)
 = \,
 2^{\frac{n-1}{2}}\,
 \big( \hspace{-.05cm} \begin{array}{cc}  1 & 1  \end{array} \hspace{-.05cm} \big)
 \prod_{k=n-1}^1
 \widehat T(\delta_k) \,
 \bigg( \hspace{-.1cm} \begin{array}{c} 
  0 \\ 1
  \end{array} \hspace{-.1cm} \bigg).
 \nonumber
\eea
In order to compute the matrix product in (\ref{Zn pm}), it is convenient to perform a change of basis which diagonalises $\widehat T(0)$, namely
\begin{equation}
\label{T U basis}
 T(\delta) = U^\dag \, \widehat T(\delta) \, U ,
  \qquad\quad 
  U = \frac{1}{2}
 \begin{pmatrix}
  1+\iu & -(1+\iu) \\
  1-\iu & 1-\iu
 \end{pmatrix} .
\end{equation}
From (\ref{hat T def}) and (\ref{T U basis}) we can explicitly write the transfer matrix in the new basis
\begin{equation}
 T(0) = \frac{1}{\sqrt{2}}
 \begin{pmatrix}
  1+\iu & 0 \\
  0 & 1-\iu
 \end{pmatrix} ,
 \qquad\quad
 T(1/2) = \frac{1}{\sqrt{2}}
 \begin{pmatrix}
  0   & 1-\iu \\
  1+\iu & 0
 \end{pmatrix} ,
\end{equation}
and therefore the partition function (\ref{Zn pm}) can be rewritten as follows
\begin{equation} \label{eq:transfer matrix}
 Z_n = 2^{\frac{n}{2}-1}\,
  \big( \hspace{-.05cm} \begin{array}{cc}  1 & 1  \end{array} \hspace{-.05cm} \big)\;
 T(0)
 \prod_{k=n-1}^1 T(\delta_k)
 \bigg( \hspace{-.1cm} \begin{array}{c} 
  1 \\ 1
  \end{array} \hspace{-.1cm} \bigg).
\end{equation}
Now, it is convenient to move all $T(0)$'s in the product of \eqref{eq:transfer matrix} to the left of all the $T(1/2)$'s.
To do this, one observes that $T(0)T(1/2) = T(1/2)T(0)^{-1}$ and $T(1/2)T(0) = T(0)^{-1}T(1/2)$. 
By employing the latter rule, one finds 
\begin{equation}
\label{Zn final step}
 Z_n = 2^{\frac{n}{2}-1}\,
  \big( \hspace{-.05cm} \begin{array}{cc}  1 & 1  \end{array} \hspace{-.05cm} \big)\;
 T(0)^{1+\sum_i(1-2\delta_i)(-1)^{\sum_{j=i}^{n-1}2\delta_j}}\,
 T(1/2)^{\sum_i 2\delta_i}
 \bigg( \hspace{-.1cm} \begin{array}{c} 
  1 \\ 1
  \end{array} \hspace{-.1cm} \bigg).
\end{equation}
The factor $1-2\delta_i$ within the sum in the exponent of $T(0)$ selects only the sites where $\delta_i = 0$, while the other factor $(-1)^{\sum_{j=i}^{n-1}2\delta_j}$ counts all the $T(1/2)$'s on the left of site $i$.
The exponent of $T(0)$ can be rewritten as
\begin{equation}
 s(\boldsymbol\delta) = 1 + \sum_{i=1}^{n-1}(1-2\delta_i)(-1)^{\sum_{j=i}^{n-1}2\delta_j} = 1+ \frac{1-(-1)^{\sum_i2\delta_i}}{2} + \sum_{i=1}^{n-1}(-1)^{\sum_{j=i}^{n-1}2\delta_j} .
\end{equation}
Since $T(0)$ is diagonal, its powers can be easily performed.
Moreover, since $T(1/2)^2=\mathbb{I}_2$, every integer power of $T(1/2)$ is simply $\mathbb{I}_2$ if the exponent is even, and $T(1/2)$ otherwise. Thus, we have
\bea
 T(0)^{s(\boldsymbol\delta)} &=&
  \begin{pmatrix}
  e^{\frac{\iu\pi}{4}s(\boldsymbol\delta)} & 0 \\
  0   & e^{-\frac{\iu\pi}{4}s(\boldsymbol\delta)}
 \end{pmatrix} ,
\\
\rule{0pt}{.9cm}
 T(1/2)^{2\sum_i\delta_i} &=&
  \begin{pmatrix}
  [1+(-1)^{\sum_i2\delta_i}]/2 & e^{-\frac{i\pi}{4}}[1-(-1)^{\sum_i2\delta_i}]/2  \\
  \rule{0pt}{.5cm}
  e^{\frac{i\pi}{4}} [1-(-1)^{\sum_i2\delta_i}]/2 & [1+(-1)^{\sum_i2\delta_i}]/2
 \end{pmatrix} .
\eea
Finally, (\ref{Zn final step}) tells us that we just need to multiply this two matrices and to sum the four elements of the resulting matrix.
After some of algebra, we get 
\begin{equation} 
\label{eq:q_n from Ising}
 m_n(\boldsymbol\delta) = 2^{n/2} \bigg[  \frac{1+(-1)^{\sum_i2\delta_i}}{2}\cos\left(\frac{\pi}{4} s(\boldsymbol\delta) \right) + 
 \frac{1-(-1)^{\sum_i2\delta_i}}{2}\cos\left(\frac{\pi}{4} \left(s(\boldsymbol\delta)-1\right) \right)  \bigg] .
\end{equation}
By inspection of the two cases of $\sum_i2\delta_i$ even or odd, it is clear that (\ref{eq:q_n from Ising}) 
equals \eqref{eq:r from EZ + correspondence}.

\end{appendices}



\section*{References}


\begin{thebibliography}{99}


\bibitem{rev}
L. Amico, R. Fazio, A. Osterloh, and V. Vedral, 
Rev. Mod. Phys. {\bf 80}, 517 (2008);\\
J. Eisert, M. Cramer, and M. B. Plenio, 
Rev. Mod. Phys. {\bf 82}, 277 (2010);\\
P. Calabrese, J. Cardy, and B. Doyon Eds, J. Phys. A {\bf 42} 500301 (2009).




\bibitem{partial}
A. Peres, 
Phys. Rev. Lett.\  {\bf 77},  1413 (1996);\\
K. Zyczkowski, P. Horodecki, A. Sanpera and M. Lewenstein,
Phys. Rev. A {\bf 58}, 883 (1998);\\
J. Eisert and M. B. Plenio, 
 J. Mod. Opt.  {\bf 46}, 145 (1999). 

\bibitem{vw-02}
G. Vidal and R. F. Werner, 
Phys. Rev. A {\bf 65}, 032314 (2002).


\bibitem{neg-mon}
J. Eisert, 
quant-ph/0610253;\\
M. B. Plenio,
Phys. Rev. Lett.\  {\bf 95},  090503 (2005).



\bibitem{cct-neg-letter}
P. Calabrese, J. Cardy, and E. Tonni, 
Phys. Rev. Lett. {\bf 109}, 130502 (2012).

\bibitem{cct-neg-long}
P. Calabrese, J. Cardy, and E. Tonni, 
J. Stat. Mech.  P02008 (2013).

\bibitem{ctt-13}
P. Calabrese, L. Tagliacozzo and E. Tonni, 
J. Stat. Mech. P05002 (2013).

\bibitem{rr-14}
M. Rangamani and M. Rota, 
JHEP 1410 (2014) 60;\\ 
M. Kulaxizi, A. Parnachev, and G. Policastro, 
JHEP 1409 (2014) 010;\\
E. Perlmutter, M. Rangamani, and M. Rota, arXiv:1506.01679. 


\bibitem{cct-neg-T}
P. Calabrese, J. Cardy, and E. Tonni, 
J. Phys. A {\bf 48},  015006 (2015).




\bibitem{ez-14}
V. Eisler and Z. Zimboras, 
New J. Phys. {\bf 16}, 123020  (2014).

\bibitem{ctc-14}
A. Coser, E. Tonni and P. Calabrese, 
J. Stat. Mech. P12017 (2014).

\bibitem{hd-15}
M. Hoogeveen and B. Doyon,
Nucl.~Phys.~B~{\bf 898}, 78 (2015).

\bibitem{wcr-15}
X. Wen, P.-Y. Chang, and S. Ryu,
arXiv:1501.00568.

\bibitem{c-13} C. Castelnovo, 
Phys. Rev. A {\bf 88}, 042319 (2013);\\
C. Castelnovo, Phys. Rev. A {\bf 89}, 042333 (2014).

\bibitem{lv-13}
Y. A. Lee and G. Vidal, 
Phys. Rev. A {\bf 88}, 042318 (2013).


\bibitem{kor}
R. Santos, V. Korepin, and S. Bose,
Phys. Rev. A {\bf 84}, 062307 (2011);\\
R. Santos and V. Korepin,
J. Phys. A {\bf 45}, 125307 (2012).

\bibitem{ahjk-14}
C. M. Ag\'on, M. Headrick, D. L. Jafferis, and S. Kasko, 
Phys. Rev. D {\bf 89}, 025018 (2014).

\bibitem{dct-15}
C. De Nobili, A. Coser and E. Tonni, 
J.~Stat.~Mech. P06021 (2015).


\bibitem{a-13}
V. Alba, 
J. Stat. Mech. P05013 (2013).

\bibitem{AlbaLauchlin-neg}
C. Chung, V. Alba, L. Bonnes, P. Chen, and A. Lauchli,
Phys. Rev. B {\bf 90}, 064401 (2014).

\bibitem{wmb-09}
 H. Wichterich, J. Molina-Vilaplana and S. Bose,
 Phys. Rev. A {\bf 80}, 010304 (2009).

\bibitem{aw-08}
J. Anders and A. Winter,
Quantum Inf. Comput. {\bf 8}, 0245  (2008);\\
J. Anders, 
Phys. Rev. A {\bf 77}, 062102 (2008).

\bibitem{fcga-08}
A. Ferraro, D. Cavalcanti, A. Garcia-Saez, and A. Acin,
Phys. Rev. Lett. {\bf 100}, 080502 (2008);\\
D. Cavalcanti, A. Ferraro, A. Garcia-Saez, and A. Acin,
Phys. Rev. A {\bf 78}, 012335 (2008).

\bibitem{Neg3}
H. Wichterich, J. Vidal, and S. Bose,  
Phys. Rev. A {\bf 81}, 032311 (2010).

\bibitem{sod}
A. Bayat, P. Sodano, and S. Bose,
Phys. Rev. Lett. {\bf 105}, 187204 (2010);\\
A. Bayat, P. Sodano, and S. Bose,
Phys. Rev. B {\bf 81}, 064429 (2010);\\
P. Sodano, A. Bayat, and S. Bose,
Phys. Rev. B {\bf 81}, 100412 (2010);\\
A. Bayat, S. Bose, P. Sodano, and H. Johannesson,
Phys. Rev. Lett. {\bf 109}, 066403 (2012).



\bibitem{ez-15}
V. Eisler and Z. Zimboras, 
New~J.~Phys.~{\bf 17}, 053048 (2015).


\bibitem{ip-10}
F. Igloi and I. Peschel, 
EPL {\bf 89},  40001 (2010).

\bibitem{atc-10}
V. Alba, L. Tagliacozzo, and P. Calabrese, 
Phys. Rev. B {\bf 81} 060411  (2010).


\bibitem{vlrk-03}
G. Vidal, J. Latorre, E. Rico and A. Kitaev,
Phys. Rev. Lett. {\bf 90}, 227902 (2003);\\
J. I. Latorre, E. Rico, and G. Vidal,
Quant. Inf. Comp. {\bf 4}, 048 (2004).

\bibitem{kleinert}
H.~Kleinert,
{\it Path Integrals in Quantum Mechanics, Statistics, Polymer Physics, and Financial Markets},
World Scientific, 5th ed., 2009.

\bibitem{Holzhey} 
C. Holzhey, F. Larsen, and F. Wilczek,
Nucl. Phys. B {\bf 424}, 443 (1994);\\
C. G. Callan and F. Wilczek, 
Phys. Lett. B {\bf 333}, 55  (1994).

\bibitem{cc-04}
P. Calabrese and J. Cardy,
J. Stat. Mech. P06002 (2004).

\bibitem{cc-rev}
P. Calabrese and J. Cardy,
J. Phys. A {\bf 42}, 504005 (2009).

\bibitem{fcm-10}
M Fagotti, P Calabrese, and J E Moore,
Phys. Rev. B {\bf 83}, 045110 (2011).

\bibitem{ccd-09}
J. Cardy,  O. Castro-Alvaredo, and B. Doyon,
J. Stat. Phys. {\bf 130},  129 (2008).

\bibitem{cl-08}
P. Calabrese and A. Lefevre,
Phys. Rev. A {\bf 78}, 032329 (2008).

\bibitem{jk-04}
B.-Q. Jin and V. E. Korepin,
J. Stat. Phys. {\bf 116}, 79 (2004).

\bibitem{cct-09}
P. Calabrese, J. Cardy, and E. Tonni, 
J. Stat. Mech. P11001 (2009).

\bibitem{cct-11}
P. Calabrese, J. Cardy, and E. Tonni, 
J. Stat. Mech. P01021 (2011).


\bibitem{cg-08}
M. Caraglio and F. Gliozzi, 
JHEP 0811: 076 (2008).


\bibitem{ch-05}
H. Casini, C. D. Fosco, and M. Huerta,
J. Stat. Mech. P05007 (2005).

\bibitem{fps-08}
S. Furukawa, V. Pasquier, and J. Shiraishi, 
Phys. Rev. Lett. {\bf 102}, 170602 (2009).

\bibitem{c-10}
P. Calabrese, J. Stat. Mech. P09013 (2010).

\bibitem{headrick}
M. Headrick,
Phys. Rev. D {\bf 82}, 126010 (2010).

\bibitem{hlr-13}
M. Headrick, A. Lawrence, and M. Roberts, 
J. Stat. Mech. P02022 (2013). 

\bibitem{ctt-14}
A. Coser, L. Tagliacozzo, and E. Tonni,
J. Stat. Mech.  P01008 (2014).



\bibitem{atc-11}
V. Alba, L. Tagliacozzo, and P. Calabrese, 
 J. Stat. Mech.  P06012 (2011).
 
\bibitem{rg-12} 
M. Rajabpour and F. Gliozzi,
J. Stat. Mech. P02016 (2012).

\bibitem{f-12}
M. Fagotti,
EPL {\bf 97}, 17007 (2012).

\bibitem{cz-13}
B. Chen and J. Zhang, 
JHEP 1311 (2013) 164.


\bibitem{RT}
S. Ryu and T. Takayanagi, 
Phys. Rev. Lett. {\bf 96},  181602 (2006);\\
S. Ryu and T. Takayanagi, 
JHEP 0608: 045 (2006);\\
T. Nishioka, S. Ryu, and T. Takayanagi, 
J. Phys. A {\bf 42},  504008 (2009).



\bibitem{hol}
V. E. Hubeny and M. Rangamani,
JHEP 0803: 006 (2008);\\
E. Tonni, 
JHEP 1105:004 (2011);\\
P. Hayden, M. Headrick and A. Maloney
Phys. Rev. D {\bf 87}, 046003 (2013);\\
T. Faulkner, 
arXiv:1303.7221;\\
T. Hartman, 
arXiv:1303.6955;\\
T. Faulkner, A. Lewkowycz, and J. Maldacena, 
JHEP 1311 (2013) 074;\\
P. Fonda, L. Giomi, A. Salvio and E. Tonni, 
JHEP 1502 (2015) 005.


\bibitem{cft-high-dims}
J. Cardy, 
J. Phys. A {\bf 46}  285402 (2013);\\
H. Casini and M. Huerta,
JHEP 0903: 048 (2009);\\
H. Casini and M. Huerta,
Class. Quant. Grav. {\bf 26}, 185005 (2009);\\
N. Shiba, 
JHEP 1207:100 (2012);\\
L.-Y. Hung, R. C. Myers, and M.  Smolkin, 
JHEP 1410 (2014) 178;\\
H. Schnitzer,
arXiv:1406.1161;\\
C. A. Agon, I. Cohen-Abbo, and H. J. Schnitzer, arxiv:1505.03757.





\bibitem{theta books}
D. Mumford,
{\it Tata lectures on Theta III}, Progress in Mathematics {\bf 97},
Birkh\"auser, Boston 1991;\\
J. Igusa,
{\it Theta Functions}, Springer-Verlag (1972).


\bibitem{Fay book}
J. Fay, {\it Theta functions on Riemann surfaces}, Lecture Notes in Mathematics {\bf 352}, Springer-Verlag, 1973.

\bibitem{bosonization higher genus}
L. J. Dixon, D. Friedan, E. J. Martinec and S. H. Shenker, 
Nucl. Phys. B {\bf 282},  13 (1987);\\
Al. B. Zamolodchikov, 
Nucl. Phys. B {\bf 285},  481 (1987);\\ 
L. Alvarez-Gaum\'e, G. W. Moore, and C. Vafa,
Commun. Math. Phys. {\bf 106}, 1 (1986);\\
E. Verlinde and H. Verlinde,
Nucl. Phys. B {\bf 288},  357 (1987); \\
V. G. Knizhnik, 
Commun. Math. Phys.  {\bf 112},  567 (1987);\\
M. Bershadsky and A. Radul,
Int. J. Mod. Phys. A {\bf 2}, 165 (1987).

\bibitem{dvv}
R. Dijkgraaf, E. P. Verlinde, and H. L. Verlinde,
Commun. Math. Phys.  {\bf 115},  649 (1988).

\bibitem{ginsparg}
P.~Ginsparg,
{\it Applied Conformal Field theory},
Les Houches lecture notes (1988),
arXiv:hep-th/9108028.


\bibitem{cct-15}
A. Coser, E. Tonni, and P. Calabrese,
arXiv:1503.09114.

\bibitem{fc-10}
M. Fagotti and P. Calabrese, 
J. Stat. Mech. P04016 (2010).

\bibitem{aepw-02}
K. Audenaert, J. Eisert, M. B. Plenio, and R. F. Werner,  
Phys. Rev. A {\bf 66}, 042327 (2002).

\bibitem{br-04}
A. Botero and B. Reznik,  
Phys. Rev. A {\bf 70}, 052329 (2004).

\bibitem{pe-09}
I. Peschel and V. Eisler,
J. Phys. A {\bf 42},  504003 (2009).

\bibitem{mrpr-09}
S. Marcovitch, A. Retzker, M. B. Plenio and B. Reznik,
Phys. Rev. A {\bf 80}, 012325 (2009).




\bibitem{cc-10}
J. Cardy and P. Calabrese, J. Stat. Mech. P04023 (2010).

\bibitem{ccen-10}
P. Calabrese, M. Campostrini, F. Essler and B. Nienhius,
Phys. Rev. Lett {\bf 104}, 095701 (2010).

\bibitem{ce-10}
P. Calabrese and F. H. L. Essler,
J. Stat. Mech.  P08029 (2010).

\bibitem{un-vari}
J. C. Xavier and F. C. Alcaraz, 
Phys. Rev. B {\bf 83},  214425 (2011);\\
M. Fagotti and P. Calabrese,
J. Stat. Mech. P01017 (2011);\\
M. Dalmonte, E. Ercolessi, L. Taddia,
Phys. Rev. B {\bf  84}, 085110 (2011); \\
M. Dalmonte, E. Ercolessi, L. Taddia,
Phys. Rev. B {\bf 85}, 165112 (2012);\\
P. Calabrese, M. Mintchev, and E. Vicari,
J. Stat. Mech. P09028 (2011).

\bibitem{pascazio-08}
P. Facchi, G. Florio, C. Invernizzi and S. Pascazio, 
Phys. Rev. A {\bf 78}, 052302 (2008).

\bibitem{cct-prep}
A. Coser, E. Tonni, P. Calabrese, in preparation. 

\bibitem{eg-03}
V. Enolski and T. Grava,
Int. Math. Res. Not. {\bf 32}, (2004) 1619.


\bibitem{bateman}
A. Erderlyi,
{\it Higher transcendental functions}, vol. I, Mc Graw Hill, (1953).

\bibitem{derrida}
B.~Derrida, M.~Mend\`ez~France, and J.~Peyri\`ere,
J.~Stat.~Phys. {\bf 45}, 439 (1986).




























\end{thebibliography}
\end{document}